\documentclass [12pt]{article}
\usepackage{amsmath,amssymb,cite}

\usepackage{tabularx}
\usepackage[english]{babel}
\usepackage[dvips]{graphicx}
\usepackage{indentfirst}
\usepackage{epsfig}
\numberwithin{equation}{section}

\begin{document}

\title{
Impact of Supersymmetry on Emergent Geometry in Yang-Mills Matrix Models II}

\author{Badis Ydri\footnote{Email:ydri@stp.dias.ie,~badis.ydri@univ-annaba.org.}\\
Institute of Physics, BM Annaba University,\\
BP 12, 23000, Annaba, Algeria.
}

\maketitle
\abstract{We present a study of $D=4$ supersymmetric Yang-Mills matrix models with $SO(3)$ mass terms based on the Monte Carlo method.

In the bosonic models we show the existence of an exotic first/second order transition from a phase with a well defined background geometry (the fuzzy sphere) to a phase with commuting matrices with no geometry in the sense of Connes. At the transition point the sphere expands abruptly to infinite size then  it  evaporates as we increase the temperature (the gauge coupling constant). The transition looks first order due to the discontinuity in the action whereas it looks second order due to the divergent peak in the specific heat.

The fuzzy sphere is stable for the supersymmetric models in the sense that the bosonic phase transition is turned into a very slow crossover transition. The transition point  is found to scale to zero with $N$. We conjecture that the transition from the background sphere to the phase of commuting matrices is associated with spontaneous supersymmetry breaking.

The eigenvalues distribution of any of the bosonic matrices in the matrix phase is found to be given by a non-polynomial law obtained from the fact that the joint probability distribution of the four matrices is uniform inside a solid ball with radius $R$. The eigenvalues of the gauge field on the background geometry are also found to be distributed according to this non-polynomial law.

}
\section{Introduction}
Reduced Yang-Mills theories play a central role in the nonperturbative definitions of $M$-theory and superstrings. The BFSS (Banks-Fischler-Shenker-Susskind) conjecture \cite{Banks:1996vh} relates discrete light-cone quantization (DLCQ) of $M-$theory to the theory of $N$ coincident $D0$ branes which at low energy (small velocities and/or string coupling) is the reduction to $0+1$ dimension  of the $10$ dimensional $U(N)$ supersymmetric Yang-Mills gauge theory \cite{Witten:1995im}. The BFSS model is therefore a Yang-Mills quantum mechanics  which is supposed to be the UV completion of $11$ dimensional supergravity. As it turns out the BFSS action is nothing else but the regularization of the supermembrane action in the light cone gauge \cite{deWit:1988ig}. 

The BMN model \cite{Berenstein:2002jq} is a generalization of the BFSS model to curved backgrounds. It is obtained by adding to the BFSS action a one-parameter mass deformation corresponding to the maximally supersymmetric pp-wave background of $11$ dimensional supergravity. See for example \cite{KowalskiGlikman:1984wv,Blau:2001ne,Blau:2002dy}. We note, in passing, that all maximally supersymmetric pp-wave geometries can arise as Penrose limits of $AdS_p\times S^q$ spaces \cite{penrose}.

The IKKT model \cite{Ishibashi:1996xs} is, on the other hand, a Yang-Mills matrix model obtained by dimensionally reducing $10$ dimensional $U(N)$ supersymmetric Yang-Mills gauge theory to $0+0$ dimensions.  The IKKT model is postulated to provide a constructive definition of type II B superstring theory and for this reason it is also called type IIB matrix model. Supersymmetric analogue of the IKKT model also exists in dimensions $d=3,4$ and $6$ while the partition functions converge only in dimensions $d=4,6$ \cite{Krauth:1998yu,Austing:2001pk}. 

The IKKT Yang-Mills matrix models can be thought of as continuum Eguchi-Kawai reduced models as opposed to the usual lattice Eguchi-Kawai reduced model formulated in \cite{Eguchi:1982nm}. We point out here the similarity between the conjecture that the lattice Eguchi-Kawai reduced model allows us  to recover the full gauge theory in the large $N$ theory and the conjecture that the IKKT matrix model allows us to recover type II B superstring. 

The relation between the BFSS Yang-Mills quantum mechanics and the IKKT Yang-Mills matrix model is discussed at length in the seminal paper \cite{Connes:1997cr} where it is also shown that toroidal compactification of the D-instanton action (the bosonic part of the IKKT action) yields, in a very natural way,  a noncommutative Yang-Mills theory on a dual noncommutative torus \cite{Connes:1987ue}. From the other hand, we can easily check that the ground state of the D-instanton action is given by commuting matrices which can be diagonalized simultaneously with the eigenvalues giving the coordinates of the D-branes. Thus at tree-level an ordinary spacetime emerges from the bosonic truncation of the IKKT action while higher order quantum corrections will define a noncommutative spacetime.

In summary, Yang-Mills matrix models which provide a constructive definition of string theories will naturally lead to emergent geometry \cite{Seiberg:2006wf} and non-commutative gauge theory \cite{Aoki:1999vr,Aoki:1998vn}. Furthermore, non-commutative geometry \cite{Connes:1994yd,GraciaBondia:2001tr} and their non-commutative field theories \cite{Douglas:2001ba,Szabo:2001kg} play an essential role in the non-perturbative dynamics of superstrings and $M$-theory. Thus the connections between non-commutative field theories, emergent geometry and matrix models from one side and string theory from the other side run deep. 

It seems therefore natural that Yang-Mills matrix models provide a non-perturbative framework for emergent spacetime geometry and non-commutative gauge theories. Since non-commutativity  is the only extension which preserves maximal supersymmetry, we also hope that Yang-Mills matrix models will provide a regularization which preserves supersymmetry \cite{Nishimura:2009xm}.

In this article we will explore in particular the possibility of using IKKT Yang-Mills matrix models in dimensions $4$ and $3$ to provide a non-perturbative definition of emergent spacetime geometry, non-commutative gauge theory and supersymmetry in two dimensions. From our perspective in this article, the phase of commuting matrices has no geometry in the sense of Connes and thus we need to modify the models so that a geometry with a well defined spectral triple can also emerge alongside the phase of commuting matrices. 

There are two solutions to this problem. The first solution is given by adding mass deformations which preserve supersymmetry to the flat IKKT Yang-Mills matrix models \cite{Bonelli:2002mb} or alternatively by an Eguchi-Kawai reduction of the mass deformed BFSS Yang-Mills quantum mechanics constructed in \cite{Kim:2006wg,Kim:2002cr,Hyun:2002fk,Hyun:2003se}. The second solution, which we have also considered in this article, is given by deforming the flat Yang-Mills matrix model in $D=4$ using the powerful formalism  of cohomological Yang-Mills theory  \cite{Moore:1998et,Hirano:1997ai,Kazakov:1998ji,Hoppe:1999xg}. 

These mass deformed or cohomologically deformed IKKT Yang-Mills matrix models are the analogue of the BMN model and they  typically include a Myers term \cite{Myers:1999ps} and thus they will sustain the geometry of the fuzzy sphere \cite{Hoppe:1982,Madore:1991bw} as a ground state which at large $N$ will approach the geometry of the ordinary sphere, the ordinary plane or the non-commutative plane depending on  the scaling limit. Thus a non-perturbative formulation of non-commutative gauge theory in two dimensions can be captured rigorously   within these models \cite{CarowWatamura:1998jn,Iso:2001mg,Presnajder:2003ak}. See also \cite{Ishiki:2008vf,Ishiki:2009vr}.

This can in principle be generalized to other fuzzy spaces \cite{Balachandran:2005ew} and higher dimensional non-commutative gauge theories by considering appropriate mass deformations of the flat IKKT Yang-Mills matrix models.

The problem or virtue of this construction, depending on the perspective, is that in these Yang-Mills matrix models the geometry of the fuzzy sphere collapses under quantum fluctuations  into the phase of commuting matrices. Equivalently, it is seen that the geometry of the fuzzy sphere emerges from the dynamics of a random matrix theory \cite{DelgadilloBlando:2007vx,Azuma:2004zq}. Supersymmetry is naturally expected to stabilize the spacetime geometry, and in fact the non stability of the non-supersymmetric vacuum should have come  as no surprise  to us \cite{Witten:2000zk}.  

We should mention here the approach  of \cite{Steinacker:2003sd} in which a noncommutative Yang-Mills gauge theory on the fuzzy sphere emerges also from the dynamics of a random matrix theory. The fuzzy sphere is stable in the sense that the transition to commuting matrices is pushed towards infinite gauge coupling at large $N$ \cite{O'Connor:2006wv}. This was achieved by considering a very special non-supersymmetric mass deformation which is quartic in the bosonic matrices. This construction was extended  to a  noncommutative gauge theory on the fuzzy sphere based on co-adjoint orbits in \cite{Steinacker:2007iq}. 
 
Let us also note here that the instability and the phase transition discussed here were also observed on the non-commutative torus in \cite{Bietenholz:2004wk,Bietenholz:2006cz,Bietenholz:2005iz,Azeyanagi:2008bk,Azeyanagi:2007su} where the twisted Eguchi-Kawai model was employed as a non-perturbative regularization of non-commutative Yang-Mills gauge theory  \cite{Ambjorn:2000cs,Ambjorn:1999ts,Ambjorn:2000nb}.

In this article we will then study using the Monte Carlo method the mass deformed Yang-Mills matrix model in $D=4$ as well as a particular truncation to $D=3$. We will also derive and study a one-parameter cohomological deformation of the Yang-Mills matrix model which coincides with the mass deformed model in $D=4$ when the parameter is tuned appropriately. We will show that the first/second order phase transition  from the fuzzy sphere to the phase of commuting matrices observed in the bosonic models is converted  in the supersymmetric models into a very slow crossover transition with an arbitrary small transition point in the large $N$ limit. We will determine the eigenvalues distributions for both $D=4$ and $D=3$ throughout the phase diagram. 

This article is organized as follows. In section $2$ we will derive the mass deformed Yang-Mills quantum mechanics  from the requirement of supersymmetry and then reduce it further to obtain Yang-Mills matrix model in $D=4$ dimensions. In section $3$ we will derive a one-parameter family of cohomologically deformed models and then show that the mass deformed model constructed in section $2$ can be obtained for a particular value of the parameter. In section $4$ we report our first Monte Carlo results for the model $D=4$ including the eigenvalues distributions and also comment on the $D=3$ model obtained by simply setting the fourth matrix to $0$. 
We conclude in section $5$ with a comprehensive summary of the results and discuss future directions. 

\section{Mass Deformation of $D=4$ Super Yang-Mills Matrix Model}
\subsection{Deformed Yang-Mills Quantum Mechanics in $4$D}
The ${\cal N}=1$ supersymmetric Yang-Mills theory reduced to one dimension is given by the supersymmetric Yang-Mills quantum mechanics (with $D_0={\partial}_0-i[X_0,.]$)
\begin{eqnarray}
{\cal L}_0=\frac{1}{g^2}Tr\bigg(\frac{1}{2}(D_0X_i)^2+\frac{1}{4}[X_i,X_j]^2-\frac{1}{2}\bar{\psi}{\gamma}^0D_0\psi+\frac{i}{2}\bar{\psi}{\gamma}^i[X_i,\psi] +\frac{1}{2}F^2\bigg).
\end{eqnarray}
The corresponding supersymmetric transformations are
\begin{eqnarray}
&&\delta_0 X_{0}=\bar{\epsilon}{\gamma}_{0}\psi\nonumber\\
&&\delta_0 X_{i}=\bar{\epsilon}{\gamma}_{i}\psi\nonumber\\
&&\delta_0 \psi=\bigg(-\frac{1}{2}[{\gamma}^{0},{\gamma}^{i}]D_0X_i+\frac{i}{4}[{\gamma}^{i},{\gamma}^{j}][X_i,X_j]+i{\gamma}_5F\bigg)\epsilon\nonumber\\
&&\delta_0 F=-i\bar{\epsilon}{\gamma}_5{\gamma}_{0}{D}_{0}\psi+\bar{\epsilon}{\gamma}_5{\gamma}_{i}[{X}_{i},\psi].
\end{eqnarray}
Let $\mu$ be a constant mass parameter. A mass deformation of the Lagrangian density ${\cal L}_0$ takes the form
\begin{eqnarray}
{\cal L}_{\mu}={\cal L}_0+\frac{\mu}{g^2}{\cal L}_1+\frac{{\mu}^2}{g^2}{\cal L}_2+...\label{4DL}
\end{eqnarray}
The Lagrangian density ${\cal L}_0$ has mass dimension $4$. The corrections ${\cal L}_1$ and ${\cal L}_2$ must have mass dimension $3$ and $2$ respectively. We recall that the Bosonic matrices $X_0$ and $X_{a}$ have mass dimension $1$ whereas the Fermionic matrices ${\psi}_i$ have mass dimension $\frac{3}{2}$. A typical term in the Lagrangian densities ${\cal L}_1$ and ${\cal L}_2$ will contain $n_f$ Fermion matrices, $n_b$ Boson matrices and $n_t$ covariant time derivatives. Clearly for ${\cal L}_1$ we must have $\frac{3}{2}n_f+n_b+n_t=3$. There are only three solutions $(n_f,n_b,n_t)=(2,0,0),(0,3,0),(0,2,1)$. For ${\cal L}_2$ we must have $\frac{3}{2}n_f+n_b+n_t=2$ and we have only one solution $(n_f,n_b,n_t)=(0,2,0)$. Thus the most general forms of ${\cal L}_1$ and ${\cal L}_2$ are
\begin{eqnarray}
{\cal L}_1=Tr\bigg(\bar{\psi}M{\psi}+\frac{1}{3!}S_{abc}X_aX_bX_c+J_{ab}X_aD_0X_b\bigg).
\end{eqnarray}
\begin{eqnarray}
{\cal L}_2=Tr\bigg(-\frac{1}{2!}S_{ab}X_aX_b\bigg).
\end{eqnarray}
Clearly for ${\cal L}_3$ we must have $\frac{3}{2}n_f+n_b+n_t=1$ which can not be satisfied. Thus the correction ${\cal L}_3$ and all other higher order corrections vanish identically.

We will follow the method of \cite{Kim:2006wg} to determine the exact form of the mass deformation.  We find the fermionic mass term
\begin{eqnarray}
{\cal L}_{\psi}=Tr\bar{\psi}\bigg(ia{\bf 1}_4+\frac{1}{2}H_{ij}{\gamma}^0[{\gamma}^i,{\gamma}^j]+c{\gamma}^1{\gamma}^2{\gamma}^3\bigg)\psi.
\end{eqnarray}
The numerical coefficients $a$, $H_{ij}$ and $c$ will be constrained further under the requirement of supersymmetry invariance.

Next we consider the bosonic terms. By rotational invariance we can choose $J_{ab}=0$, $S_{ab}=v\delta_{ab}$ and $S_{abc}=6ie\epsilon_{abc}$ for some numerical coefficients $v$ and $e$.

The mass deformed supersymmetric transformations will be taken such that on bosonic fields they will coincide with the non deformed supersymmetric transformations so that the Fierz identity can still be used. The mass deformed supersymmetric transformations on fermionic fields will be different from  the non deformed supersymmetric transformations with a time dependent parameter ${\epsilon}\equiv {\epsilon}(t)$ which satisfies ${\partial}_0{\epsilon}=\mu \Pi{\epsilon}$. We will suppose the supersymmetric transformations
\begin{eqnarray}
&&{\delta}_{\mu} X_{0}={\delta}_0 X_{0}\nonumber\\
&&{\delta}_{\mu} X_{i}={\delta}_0 X_{i}\nonumber\\
&&{\delta}_{\mu} {\psi}={\delta}_0 {\psi}+\mu{\Delta}{\epsilon}.
\end{eqnarray}
By requiring that the  Lagrangian density (\ref{4DL}) is invariant under these transformations we can determine  precisely the form of the  mass deformed Lagrangian density  and the mass deformed supersymmetry transformations. 

A long calculation will yield  $H_{ij}=0$, $c=-3e/4$, $v=e^2-16a^2/9$, $\Pi=-(2c+e)\gamma^0\gamma^1\gamma^2\gamma^3$, $\Delta=-(4ia/3+e\gamma^1\gamma^2\gamma^3)\gamma^iX_i$. The mass deformed  Lagrangian density  and mass deformed supersymmetry transformations are given respectively by (with $ia\mu={{\mu}_1}/{4}$ and $-{3e\mu}/{4}={{\mu}_2}/{4}$) 
\begin{eqnarray}
{\cal L}_{\mu}
&=&{\cal L}_{0}+\frac{1}{4g^2}Tr\bar{\psi}\big({\mu}_1+{\mu}_2{\gamma}^1{\gamma}^2{\gamma}^3\big){\psi}-i{\epsilon}_{ijk}\frac{{\mu}_2}{3g^2}TrX_iX_jX_k-\frac{1}{18g^2}({\mu}_1^2+{\mu}_2^2)TrX_i^2.\nonumber\\
\end{eqnarray}
\begin{eqnarray}
&&{\delta}_{\mu} X_{0}=\bar{\epsilon}{\gamma}_{0}\psi\nonumber\\
&&{\delta}_{\mu} X_{i}=\bar{\epsilon}{\gamma}_{i}\psi\nonumber\\
&&{\delta}_{\mu} {\psi}=\bigg(-\frac{1}{2}[{\gamma}^{0},{\gamma}^{i}]D_0X_i+\frac{i}{4}[{\gamma}^{i},{\gamma}^{j}][X_i,X_j]-\frac{1}{3}\big({\mu}_1-{\mu}_2{\gamma}^1{\gamma}^2{\gamma}^3\big){\gamma}^iX_i\bigg)\epsilon.
\end{eqnarray}
\begin{eqnarray}
\epsilon\equiv \epsilon(t)=e^{\frac{1}{6}\big({\mu}_1{\gamma}^0-{\mu}_2{\gamma}^0{\gamma}^1{\gamma}^2{\gamma}^3\big)t}.
\end{eqnarray}
We verify that $({\delta}_{\mu} X_{\mu})^+={\delta}_{\mu} X_{\mu}$ and hence the Hermitian matrices $X_{\mu}$ remains Hermitian under supersymmetry. The corresponding supersymmetric algebra is $su(2|1)$ \cite{Kim:2006wg}.

\subsection{Truncation to $0$ Dimension}

We consider now the Lagrangian density (action) given by
\begin{eqnarray}
{\cal L}_{\mu}&=&{\cal L}_{0}+\frac{a}{4g^2}Tr\bar{\psi}\big({\mu}_1+{\mu}_2{\gamma}^1{\gamma}^2{\gamma}^3\big){\psi}-i{\epsilon}_{ijk}\frac{b{\mu}_2}{3g^2}TrX_iX_jX_k-\frac{c}{18g^2}({\mu}_1^2+{\mu}_2^2)TrX_i^2.\nonumber\\
\end{eqnarray}
\begin{eqnarray}
{\cal L}_0=\frac{1}{g^2}Tr\bigg(\frac{1}{4}[X_{\mu},X_{\nu}][X^{\mu},X^{\nu}]+\frac{i}{2}\bar{\psi}{\gamma}^{\mu}[X_{\mu},\psi] \bigg).
\end{eqnarray}
In above we have allowed for the possibility that mass deformations corresponding to the reduction to zero and one dimensions can be different by including different  coefficients $a$, $b$ and $c$ in front of the fermionic mass term, the Myers term and the bosonic mass term respectively. However we will keep the mass deformed supersymmetric transformations unchanged. 

After some algebra we find that we must have $a=2/3$, $c=1$, $b=1$ and $\mu_1=0$. The model of interest is therefore 
\begin{eqnarray}
{\cal L}_{\mu}&=&\frac{1}{g^2}Tr\bigg(\frac{1}{4}[X_{\mu},X_{\nu}][X^{\mu},X^{\nu}]+\frac{i}{2}\bar{\psi}{\gamma}^{\mu}[X_{\mu},\psi] +\frac{{\mu}_2}{6}Tr\bar{\psi}{\gamma}^1{\gamma}^2{\gamma}^3{\psi}-\frac{{\mu}_2^2}{18}TrX_i^2\nonumber\\
&-&i{\epsilon}_{ijk}\frac{{\mu}_2}{3}TrX_iX_jX_k\bigg).
\end{eqnarray}
Since $\psi$ and $\epsilon$ are Majorana spinors we can rewrite them as
\begin{eqnarray}
&&\psi=\left(
\begin{array}{c}
i{\sigma}_2({\theta}^+)^T \\
{\theta}
\end{array}
\right)~,~\epsilon=\left(
\begin{array}{c}
i{\sigma}_2({\omega}^+)^T \\
{\omega}
\end{array}
\right).
\end{eqnarray}
We compute with  $X_0=iX_4$ the action
\begin{eqnarray}
{\cal L}_{\mu}
&=&\frac{1}{g^2}Tr\bigg(\frac{1}{2}[X_{4},X_{i}]^2+\frac{1}{4}\bigg([X_i,X_j]-i\frac{{\mu}_2}{3}{\epsilon}_{ijk}X_k\bigg)^2+{\theta}^+\bigg(i[X_4,..]+{\sigma}_i[X_i,..]+\frac{{\mu}_2}{3}\bigg)\theta\bigg).\nonumber\\
\end{eqnarray}
The supersymmetric transformations are
\begin{eqnarray}
&&{\delta}_{\mu}X_0=i({\omega}^+\theta-{\theta}^+\omega)\nonumber\\
&&{\delta}_{\mu}X_i=i({\theta}^+{\sigma}_i{\omega}-{\omega}^+{\sigma}_i\theta)\nonumber\\
&&{\delta}_{\mu}\theta=\bigg(-i{\sigma}_i[X_0,X_i]-\frac{1}{2}{\epsilon}_{ijk}{\sigma}_k[X_i,X_j]+\frac{i}{3}{\mu}_2{\sigma}_iX_i\bigg)\omega.
\end{eqnarray}

\section{Cohomological Approach}
\subsection{Cohomologically Deformed Supersymmetry}
 The reduction to zero dimension of the ${\cal N}=1$ supersymmetric Yang-Mills theory in four dimensions  is given by (in Euclidean signature)
\begin{eqnarray}
S
&=&-\frac{1}{4}Tr[X_{\mu},X_{\nu}]^2-Tr{\theta}^+\bigg(i[X_4,..]+{\sigma}_a[X_a,..]\bigg)\theta+2Tr B^2.
\end{eqnarray}
The supersymmetric transformations become
\begin{eqnarray}
&&{\delta}X^{\mu}=i\bar{\omega}\bar{\sigma}^{\mu}\theta-i\bar{\theta}\bar{\sigma}^{\mu}\omega\nonumber\\
&&\delta\theta=i{\sigma}^{\mu\nu}[X^{\mu},X^{\nu}]\omega-2B\omega\nonumber\\
&&\delta\bar{\theta}=-i\bar{\omega}{\sigma}^{\mu\nu}[X^{\mu},X^{\nu}]+2B\bar{\omega}\nonumber\\
&&  \delta B=\frac{1}{2}\bar{\omega}\bar{\sigma}^{\mu}[X^{\mu},\theta]+\frac{1}{2}[X^{\mu},\bar{\theta}]\bar{\sigma}^{\mu}\omega.
\end{eqnarray}
Let us note that since we are in Euclidean signature the transformation law of $X_4$ is antihermitian rather than hermitian.

By using a contour shifting argument for the Gaussian integral over $B$ we can rewrite the auxiliary field $B$ as
\begin{eqnarray}
B=H+\frac{1}{2}[X_1,X_2].
\end{eqnarray}
We will also introduce
\begin{eqnarray}
\theta_1=\eta_2+i\eta_1~,~\theta_2=\chi_1+i\chi_2.
\end{eqnarray}
\begin{eqnarray}
\phi=\frac{1}{2}(X_3+iX_4)~,~\bar{\phi}=-\frac{1}{2}(X_3-iX_4).
\end{eqnarray}
We compute
\begin{eqnarray}
\frac{S}{2}&=&S_{\rm cohom}\nonumber\\
&=&Tr\bigg(H^2+H[X_1,X_2]+[X_i,\phi][X_i,\bar{\phi}]+[\phi,\bar{\phi}]^2-\eta_i[\phi,\eta_i]-\chi_i[\bar{\phi},\chi_i]-\eta_1\epsilon^{ij}[{\chi}_i,X_j]+\eta_2[\chi_i,X_i]\bigg).\nonumber\\
\end{eqnarray}
We have four independent real supersymmetries generated by the four independent grassmannian parameters $\xi_i$, $\rho_i$ defined by the equations $\omega_1=\xi_2+i\xi_1$ and $\omega_2=\rho_1+i\rho_2$. We look at the supercharge $Q_{1R}$ associated with $\xi_2$. We define the exterior derivative $d$ on bosons by $dB=i[Q_{1R},B]$ and on fermions by $dF=i\{Q_{1R},F\}$. The corresponding supersymmetric transformations are precisely given by $\hat{\delta}B=2\xi_2 dB$ and  $\hat{\delta}F=2\xi_2 dF$ where
\begin{eqnarray}
d X_i=\chi_i.
\end{eqnarray}
\begin{eqnarray}
d\phi=0~,~d\bar{\phi}=-\eta_2.
\end{eqnarray}
\begin{eqnarray}
  dH&=&[\phi,\eta_1].
\end{eqnarray}
\begin{eqnarray}
  d \eta_1=H~,~d\eta_2=[\bar{\phi},\phi].
\end{eqnarray}
\begin{eqnarray}
  d\chi_i&=&[\phi,X_i].
\end{eqnarray}
From these transformation laws we can immediately deduce that for any operator ${\cal O}$ we must have
\begin{eqnarray}
d^2{\cal O}=[\phi,{\cal O}]. 
\end{eqnarray}
Thus  $d^2$ is a gauge transformation generated by $\phi$ and as a consequence it is nilpotent on gauge invariant quantities such as the action.

Next we compute
\begin{eqnarray}
dTrQ=S_{\rm cohom}.
\end{eqnarray}
\begin{eqnarray}
Q=-\chi_i[X_i,\bar{\phi}]+\eta_1[X_1,X_2]+\eta_1H-\eta_2[\phi,\bar{\phi}].\label{Q}
\end{eqnarray}
Thus we have
\begin{eqnarray}
d^2TrQ= dS_{\rm cohom}=0.
\end{eqnarray}
We consider now the deformed action and deformed exterior derivative given by 
\begin{eqnarray}
S_{\rm def}=S_{\rm cohom}+\hat{S}=S_{\rm cohom}+\epsilon_1 S_1+\epsilon_2 S_2+...
\end{eqnarray}
\begin{eqnarray}
d_{\rm def}=d+\epsilon T.
\end{eqnarray}
Supersymmetric invariance requires
\begin{eqnarray}
d_{\rm def}S_{\rm def}=0.\label{supersymmetric}
\end{eqnarray}
The fact that $d^2$ is equal $0$ on gauge invariant quantities, i.e. $d^2S_{\rm cohom}=d^2S_i=0$ leads to $d^2S_{\rm def}=0$. We have the identity
\begin{eqnarray}
d_{\rm def}^2S_{\rm def}=0.
\end{eqnarray}
This is equivalent to
\begin{eqnarray}
\{d,T\}S_{\rm cohom}+\epsilon T^2S_{\rm cohom}+\epsilon\{d,T\}\hat{S}+\epsilon T^2\hat{S}=0.
\end{eqnarray}
Thus we must have among other things
\begin{eqnarray}
\{d,T\}S_{\rm cohom}=\{d,T\}\hat{S}=0.
\end{eqnarray}
In other words $\{d,T\}$ generates one of the continuous bosonic symmetries of the action $S_{\rm def}$ which are gauge transformations and the remaining rotations given by the $SO(2)$ subgroup of $SO(4)$. Following \cite{Kazakov:1998ji} we choose $\{d,T\}$ to be the rotation ${U}$ defined by
\begin{eqnarray}
{U}:X_a\longrightarrow i\epsilon_{ab}X_b~,~{\chi}_a\longrightarrow i\epsilon_{ab}{\chi}_b.
\end{eqnarray}
We have then
\begin{eqnarray}
\{d,T\}={ U}.
\end{eqnarray}
The symmetry $T$ must also satisfy
\begin{eqnarray}
T^2=0.
\end{eqnarray}
By following the method of \cite{Austing:2001ib} we can determine precisely the form of the correction $T$ from the two requirements $T^2=0$ and $\{d,T\}=U$ and also from the assumption that $T$ is linear in the fields. A straightforward calculation shows that there are two solutions but we will only consider here the one which  generates mass terms for all the bosonic fields. This is given explicitly by

\begin{eqnarray}
&&TX_i=0~,~T\chi_i=i\epsilon_{ij}X_j~,~T\phi=0\nonumber\\
&&TH=i\gamma \eta_2~,~T\eta_2=0~,~T\eta_1=-i\lambda\phi+i\gamma\bar{\phi}~,~T\bar{\phi}=0.
\end{eqnarray}
The cohomologically deformed supersymmetric transformations are therefore given by
\begin{eqnarray}
d_{\rm def} X_i=\chi_i.
\end{eqnarray}
\begin{eqnarray}
d_{\rm def}\phi=0~,~d_{\rm def}\bar{\phi}=-\eta_2.
\end{eqnarray}
\begin{eqnarray}
  d_{\rm def}H&=&[\phi,\eta_1]+i\epsilon \gamma\eta_2.
\end{eqnarray}
\begin{eqnarray}
  d_{\rm def} \eta_1=H+\epsilon(-i\lambda\phi+i\gamma\bar{\phi})~,~d_{\rm def}\eta_2=[\bar{\phi},\phi].
\end{eqnarray}
\begin{eqnarray}
  d_{\rm def}\chi_i&=&[\phi,X_i]+i\epsilon \epsilon_{ij}X_j.
\end{eqnarray}
\subsection{Cohomologically Deformed Action}

Next we need to solve the condition (\ref{supersymmetric}). The deformed action is a trace over some polynomial $P$. In the non-deformed case we have $S=dQ$ where $Q$ is a $U-$invariant expression given by (\ref{Q}). We assume that the deformed action $S_{\rm def}=TrP$ is also $U-$invariant. By using the theorem of Austing \cite{Austing:2001ib} we can conclude that the general solution of the  condition (\ref{supersymmetric}), or equivalently of the equation $d_{\rm def}TrP=0$, is
\begin{eqnarray}
S_{\rm def}=d_{\rm def}TrQ_{\rm def}+TrR_3(\phi).
\end{eqnarray}
For $SU(N)$ gauge group this result holds as long as the degree of $P$ is less than $2N/3$. Clearly when the deformation is sent to zero $d_{\rm def}\longrightarrow d$, $Q_{\rm def}\longrightarrow Q$ and $R\longrightarrow 0$. Thus we take

\begin{eqnarray}
Q_{\rm def}=Q-iR~,~R=\kappa_1R_1+\kappa_2R_2.
\end{eqnarray}
We choose $R_1$ and $R_2$ to be the $U-$invariant quantities given by
\begin{eqnarray}
R_1=\frac{1}{2}\epsilon_{ab}\chi_aX_b~,~R_2=-\eta_1\bar{\phi}.
\end{eqnarray}
We choose $R_3(\phi)$ to be the $U-$invariant quantity given by
 \begin{eqnarray}
R_3(\phi)=-\rho^2\phi^2.
\end{eqnarray}
In order to remove the deformation we must take $\epsilon\longrightarrow 0$ so that $d_{\rm def}\longrightarrow d$ and  $\rho\longrightarrow 0$ so that $S_{\rm def}\longrightarrow d TrQ_{\rm def}$ and $\kappa_i\longrightarrow 0$ so that $Q_{\rm def}\longrightarrow Q$. 

We compute
\begin{eqnarray}
S_{\rm def}&=&dTrQ-idTrR+\epsilon TTrQ-i\epsilon TTrR+TrR_3(\phi).
\end{eqnarray}
The first term $S_{\rm cohom}=dTr Q$ is the original action. We will choose the parameters so that the total action enjoys $SO(3)$ covariance with a Myers (Chern-Simons) term and mass terms for all the bosonic and fermionic matrices. We find after some algebra
\begin{eqnarray}
\gamma=\frac{\kappa_1+\kappa_2}{\epsilon}~,~\lambda=4-\frac{\kappa_1}{\epsilon}~,~\rho^2=2\epsilon(2\epsilon-\kappa_1).
\end{eqnarray}
Thus the total action becomes
\begin{eqnarray}
S_{\rm def}&=&S_{\rm cohom}+\Delta{S}_{\rm cohom}.
\end{eqnarray}
\begin{eqnarray}
\Delta{S}_{\rm cohom}&=&i\kappa_1Tr(\chi_1\chi_2-\eta_1\eta_2)+\frac{1}{2}\epsilon\kappa_1Tr X_a^2-\frac{1}{8}(\kappa_2-\epsilon\gamma)(\kappa_2-\epsilon\gamma+\epsilon\lambda)TrX_4^2\nonumber\\
&-&\frac{i}{4}(\epsilon\lambda+\epsilon\gamma+2\kappa_1 -\kappa_2+4\epsilon)Tr X_3[X_1,X_2]\nonumber\\
&=&\frac{\kappa_1}{2}Tr\theta^+\theta+\frac{1}{2}\epsilon\kappa_1Tr X_a^2+\frac{1}{4}\kappa_1(2\epsilon-\kappa_1)TrX_4^2-\frac{i}{6}(4\epsilon+\kappa_1)\epsilon_{abc}TrX_aX_bX_c.\nonumber\\
\end{eqnarray}
We introduce now 
\begin{eqnarray}
-(4\epsilon+\kappa_1)=\alpha~,~\kappa_1=-\frac{\alpha}{3}+4\zeta_0\alpha.\label{alpha}
\end{eqnarray}
Also we perform the scaling
\begin{eqnarray}
X_{\mu}\longrightarrow (2N)^{\frac{1}{4}}X_{\mu}~,~\theta\longrightarrow \sqrt{\frac{2}{N\alpha}}\frac{1}{(2N)^{\frac{1}{8}}}\theta,
\end{eqnarray}
and
\begin{eqnarray}
\alpha\longrightarrow 2(2N)^{\frac{1}{4}}\alpha.\label{scaling}
\end{eqnarray}
We get then the one-parameter family of actions given by (we set $B=0$)

\begin{eqnarray}
S_{\rm def}&=&-\frac{N}{4} Tr[X_{\mu},X_{\nu}]^2+N\frac{2i\alpha}{3}\epsilon_{abc}TrX_aX_bX_c+\frac{2N\alpha^2}{9}(1+6\zeta_0)(1-12\zeta_0)Tr X_a^2\nonumber\\
&+&4N\alpha^2\zeta_0(1-12\zeta_0)TrX_4^2-\frac{1}{N\alpha}Tr{\theta}^+\bigg(i[X_4,..]+{\sigma}_a[{X}_a,..]+\frac{2\alpha}{3}(1-12\zeta_0)\bigg)\theta.\nonumber\\
\end{eqnarray}
For stability the parameter $\zeta_0$ must be in the range
\begin{eqnarray}
0<\zeta_0<1/12.
\end{eqnarray}
This action for $\zeta_0=0$ is precisely the mass deformed action derived in section $2$. The value $\zeta_0=1/12$ will also be of interest to us in this article. This one-parameter family of actions preserves only half of the ${\cal N}=1$ supersymmetry in the sense that we can construct only two mass deformed supercharges \cite{Austing:2001ib}.

\section{Simulation Results for $D=4$ Yang-Mills Matrix Models}
\subsection{Models, Supersymmetry and Fuzzy Sphere}
We are interested in the cohomologically deformed Yang-Mills matrix models
\begin{eqnarray}
S&=&N Tr\bigg[-\frac{1}{4}[X_{\mu},X_{\nu}]^2+\frac{2i\alpha}{3}{\epsilon}_{abc}X_aX_bX_c\bigg]+N{\beta} TrX_a^2+N\beta_4 Tr X_4^2\nonumber\\
&-&\frac{1}{N\alpha}Tr{\theta}^+\bigg(i[X_4,..]+{\sigma}_a[{X}_a,..]+{\zeta}\bigg)\theta.\label{model0}
\end{eqnarray}
The range of the parameters is
 \begin{eqnarray}
\beta=\frac{2}{9}(\alpha+6\xi_0)(\alpha-12\xi_0)~,~\beta_4=4\xi_0(\alpha-12\xi_0)~,~\zeta=\frac{2}{3}(\alpha-12\xi_0).
\end{eqnarray}
\begin{eqnarray}
0\leq \xi_0\leq \frac{\alpha}{12}.
\end{eqnarray}
This action preserves two supercharges compared to the four supercharges of the original non deformed Yang-Mills matrix model \cite{Austing:2001ib}. We will be mainly interested in the "minimally" deformed Yang-Mills matrix model corresponding to the value $\xi_0=\alpha/12$ for which we have
 \begin{eqnarray}
S&=&N Tr\bigg[-\frac{1}{4}[X_{\mu},X_{\nu}]^2+\frac{2i\alpha}{3}{\epsilon}_{abc}X_aX_bX_c\bigg]-\frac{1}{N\alpha}Tr{\theta}^+\bigg(i[X_4,..]+{\sigma}_a[{X}_a,..]\bigg)\theta.\nonumber\\
\end{eqnarray}
The "maximally" deformed Yang-Mills matrix model corresponding to the value $\xi_0=0$ coincides precisely with the mass-deformed model in $D=4$ and as such it has a full ${\cal N}=1$ mass deformed supersymmetry besides the half ${\cal N}=1$  cohomologically deformed supersymmetry.  From this perspective this case is far more important than the previous one. However there is the issue of the convergence of the partition function which we will discuss shortly. In any case the "maximally" deformed Yang-Mills matrix model is given by the action 
\begin{eqnarray}
S&=&N Tr\bigg[-\frac{1}{4}[X_{\mu},X_{\nu}]^2+\frac{2i\alpha}{3}{\epsilon}_{abc}X_aX_bX_c\bigg]+N\frac{2\alpha^2}{9} TrX_a^2\nonumber\\
&-&\frac{1}{N\alpha}Tr{\theta}^+\bigg(i[X_4,..]+{\sigma}_a[{X}_a,..]+\frac{2}{3}{\alpha}\bigg)\theta.
\end{eqnarray}
The above two actions can also be rewritten as 
\begin{eqnarray}
S_{\rm SUSY}&=&NTr\bigg[-\frac{1}{4}[X_{\mu},X_{\nu}]^2+\frac{2i\alpha}{3}{\epsilon}_{abc}X_aX_bX_c\bigg]+N\tilde{\beta}\alpha^2TrX_a^2\nonumber\\
&-&\frac{1}{N\alpha}Tr{\theta}^+\bigg(i[X_4,..]+{\sigma}_a[X_a,..]+\alpha\tilde{{\xi}}\bigg)\theta.
\end{eqnarray}
\begin{eqnarray}
&&\tilde{\beta}=0~,~\tilde{\xi}=0~~{\rm cohomologically}~{\rm deformed}.\nonumber\\
&&\tilde{\beta}=\frac{2}{9}~,~\tilde{\xi}=\frac{2}{3}~~{\rm mass}~{\rm deformed}.
\end{eqnarray}
We remark that the bosonic part of the mass-deformed Yang-Mills matrix action can be rewritten as a complete square, viz
\begin{eqnarray}
S_B&=&N Tr\bigg[-\frac{1}{4}[X_{\mu},X_{\nu}]^2+\frac{2i\alpha}{3}{\epsilon}_{abc}X_aX_bX_c\bigg]+N\frac{2\alpha^2}{9} TrX_a^2\nonumber\\
&=&N Tr\bigg(\frac{i}{2}[X_{\mu},X_{\nu}]+\frac{\alpha}{3}{\epsilon}_{\mu\nu\lambda}X_{\lambda}\bigg)^2.
\end{eqnarray}
Clearly ${\epsilon}_{\mu\nu\lambda}=0$ if any of the indices $\mu$,$\nu$,$\lambda$ takes the value $4$. Generically the bosonic action of interest is given by
\begin{eqnarray}
S_B&=&N Tr\bigg[-\frac{1}{4}[X_{\mu},X_{\nu}]^2+\frac{2i\alpha}{3}{\epsilon}_{abc}X_aX_bX_c\bigg]+N\tilde{\beta}\alpha^2 TrX_a^2.
\end{eqnarray}
Here we allow $\tilde{\beta}$ to take on any value. The variation of the bosonic action for generic values of $\tilde{\beta}$ reads
\begin{eqnarray}
&&{\delta}S_B=NTrJ_4{\delta}X_4+NTrJ_b{\delta}X_b\nonumber\\
&&J_4=[X_a,[X_a,X_4]]~,~J_b=2\tilde{\beta}\alpha^2 X_b+i[F_{ab},X_a]+[X_4,[X_4,X_b]]~,~\nonumber\\
&&F_{ab}=i[X_a,X_b]+\alpha{\epsilon}_{abc}X_c.
\end{eqnarray}
Thus extrema of the model are given by $1)$ reducible representations $J_a$ of $SU(2)$, i.e $X_a=J_a$ and $X_4=0$ and $2)$ commuting matrices, i.e $X_{\mu}$ belong to the Cartan sub-algebra of $SU(N)$. The identity matrix corresponds to an uncoupled mode and thus we have $SU(N)$ instead of $U(N)$. Global minima are given by irreducible representations of $SU(2)$ of dimensions $N$ and $0$. Indeed we find that the configurations $X_a=\phi L_a$, $X_4=0$ solve the equations of motion with $\phi$ satisfying the cubic equation $\phi({\phi}^2-\alpha \phi+\tilde{\beta}\alpha^2)=0$. We get the solutions
\begin{eqnarray}
{\phi}_0=0~,~{\phi}_{\pm}=\alpha\frac{1\pm \sqrt{1-4\tilde{\beta}}}{2}. 
\end{eqnarray}
We can immediately see that we must have $\tilde{\beta}{\leq}{1}/{4}$ which does indeed hold for the values of interest $\tilde{\beta}=0$ and $\tilde{\beta}=2/9$. However the action at ${\phi}_{\pm}$ is given by
\begin{eqnarray}
S_B[{\phi}_{\pm}]=\frac{N^2c_2{\phi}_{\pm}^2}{2}\alpha^2\bigg(\tilde{\beta}-\frac{1}{6}\mp\frac{1}{6}\sqrt{1-4\tilde{\beta}}\bigg).
\end{eqnarray}
We can verify that $S_B[{\phi}_{-}]$ is always positive while $S_B[{\phi}_{+}]$ is  negative for the values of $\tilde{\beta}$ such that  $\tilde{\beta}{\leq}{2}/{9}$. Furthermore we note that $S[{\phi}_0]=0$. In other words for $\tilde{\beta}{\leq}2/9$ the global minimum of the model is the irreducible representation of $SU(2)$ of maximum dimension $N$ whereas for $\tilde{\beta}> 2/9$ the global minimum of the model is the irreducible representation of $SU(2)$ of minimum dimension $0$. 

At $\tilde{\beta}=2/9$ we get ${\phi}_+=2\alpha/3$ and $S[{\phi}_+]=0$. Thus the configuration $X_a=\frac{2\alpha}{3}L_a$ becomes degenerate with the configuration $X_a=0$. However there is an entire $SU(N)$ manifold of configurations $X_a=\frac{2\alpha}{3}UL_aU^+$ which are equivalent to the fuzzy sphere configuration. In other words the fuzzy sphere configuration is  still favored although now due to entropy. Thus there is a first order transition at $\tilde{\beta}=2/9$ when the classical ground state switches from $X_a=\frac{2}{3}L_a$ to $X_a=0$ as we increase $\tilde{\beta}$ through the critical value $\tilde{\beta}=2/9$.  The two values of interest $\tilde{\beta}=0$ and $\tilde{\beta}=2/9$ both lie in the regime where the fuzzy sphere is the stable classical ground state. 

This discussion holds also for the full bosonic model in which we include a mass term for the matrix $X_4$. Quantum correction, i.e. the inclusion of fermions, are expected to alter significantly this picture.



Towards the commutative limit we rewrite the action into the form (with $X_{\mu}=\alpha D_{\mu}$, $\tilde{\alpha}=\alpha\sqrt{N}=1/g$, $\tilde{\alpha}^4=1/g^2$ and $F_{ab}=i[D_a,D_b]+{\epsilon}_{abc}\phi D_c$)
\begin{eqnarray}
S&=&\frac{1}{4g^2N} TrF_{ab}^2-\frac{3\phi -2}{6g^2N}Tr\left[\epsilon_{abc}F_{ab}D_c+\phi D_a^2\right]+\frac{1}{g^2N}\bigg(\phi(\phi-1)+\tilde{\beta}\bigg)TrD_a^2\nonumber\\
&-&\frac{1}{2g^2N}Tr [D_a,D_4]^2-\frac{1}{N}Tr{\theta}^+ \big(i[D_4,..]+{\sigma}_a[{D}_a,..] + \tilde{\xi} \big){\theta}.
\end{eqnarray}
The $3$rd terms actually cancel for all values of $\beta$. Thus
\begin{eqnarray}
S&=&\frac{1}{4g^2N} TrF_{ab}^2-\frac{3\phi -2}{6g^2N}Tr\left[\epsilon_{abc}F_{ab}D_c+\phi D_a^2\right]-\frac{1}{2g^2N}Tr [D_a,D_4]^2\nonumber\\
&-&\frac{1}{N}Tr{\theta}^+ \big(i[D_4,..]+{\sigma}_a[{D}_a,..] + \tilde{\xi} \big){\theta}.
\end{eqnarray}
The commutative limit $N\longrightarrow \infty$ is then obvious. We write $D_a=\phi (L_a +A_a)$ and we obtain

\begin{eqnarray}
S&=&\frac{1}{4g^2} \int_{S^2} F_{ab}^2-\frac{(3\phi-2)\phi}{4g^2}{\epsilon}_{abc}\int F_{ab}A_c
-\frac{1}{2g^2}\int_{S^2}({\cal L}_aD_4)^2-\int_{S^2} {\psi}^+ \big(\phi {\sigma}_a{\cal L}_a + \tilde{\xi} \big){\psi}.\nonumber\\
\end{eqnarray}
\subsection{Path Integral, Convergence and Observables}

In the quantum theory we will integrate over $N\times N$ bosonic matrices $X_{\mu}$ and $N\times N$ fermionic matrices ${\theta}_{\alpha}^+$ and ${\theta}_{\alpha}$. The trace parts of $X_{\mu}$, ${\theta}_{\alpha}^+$ and ${\theta}_{\alpha}$ will be removed since they correspond to  free degrees of freedom. The  partition function of the model is therefore given by
\begin{eqnarray}
Z&=&\int {\cal D}X_{\mu}~{\cal D} \theta~{\cal D}{\theta}^+~{\delta}\big(Tr X_{\mu}\big)~{\delta}\big(Tr{\theta}_{\alpha}^+\big)~\delta\big(Tr{\theta}_{\alpha}\big)~e^{-S_{\rm SUSY}}\nonumber\\
&=&\int {\cal D}X_{\mu}{\delta}~\big(Tr X_{\mu}\big)~{\rm det}{\cal D}~e^{-S_B}.\label{pathI}
\end{eqnarray}
\begin{eqnarray}
{\rm det}{\cal D}&=&\int d{\theta}d{\theta}^+{\delta}(Tr{\theta}_{\alpha}){\delta}(Tr{\theta}_{\alpha}^+)e^{\frac{1}{N\alpha}Tr {\theta}^+{\cal D}{\theta}}.
\end{eqnarray}
The integration over the fermions yielded  the determinant of the $2(N^2-1)\times 2(N^2-1)$ dimensional matrix ${\cal D}=i[X_4,..]+{\sigma}_a[X_a,..]+\tilde{\xi}\alpha$. This determinant is positive definite since every eigenvalue ${\lambda}$ of ${\cal D}$ is doubly degenerate \cite{Ambjorn:2000bf}. The reason lies in the fact that the Dirac operator ${\cal D}=iX_4-iX_4^R+{\sigma}_aX_a-{\sigma}_aX_a^R+\tilde{\xi}\alpha$ is symmetric under the exchange of left and right operators, viz under $X_a\leftrightarrow -X_a^R$. A much cleaner proof goes as follows. Let $\Psi$ be an eigenstate of ${\cal D}$ with eigenvalue $\lambda$, in other words 
\begin{eqnarray}
i[X_4,\Psi]+{\sigma}_a[X_a,\Psi]+\tilde{\xi}\alpha \psi =\lambda \Psi.
\end{eqnarray}
Taking the hermitian conjugate of this equation we get
\begin{eqnarray}
i[X_4,(\Psi^+)^T]-{\sigma}_a^T[X_a,({\Psi}^+)^T]+\tilde{\xi}\alpha ({\Psi}^+)^T=\lambda ({\Psi}^+)^T.
\end{eqnarray}
In above $({\Psi}^+)^T$ is a column vector with components given by ${\Psi}^+_{1,2}$. Multiplying the above equation by ${\sigma}_2$ and defining the spinor $\tilde{\Psi}={\sigma}_2({\Psi}^+)^T$ we arrive at the equation
\begin{eqnarray}
i[X_4,\tilde{\Psi}]+{\sigma}_a[X_a,\tilde{\Psi}]+\xi\alpha \tilde{\Psi} =\lambda \tilde{\Psi}.
\end{eqnarray}
We have also used the identity ${\sigma}_a=-{\sigma}_2{\sigma}_a^T{\sigma}_2$. We conclude that $\tilde{\Psi}$ is also an eigenstate of ${\cal D}$ with the same eigenvalue $\lambda$. The spinors $\Psi$ and $\tilde{\Psi}$ are charge conjugate to each other. In above we have assumed that ${\lambda}$ is real which follows from the fact that the Dirac operator ${\cal D}=i[X_4,..]+{\sigma}_a[{X}_a,..]+\tilde{\xi}\alpha$ is hermitian. This establishes that the determinant $\det {\cal D}$ is positive definite for any configuration $X_{\mu}$ and hence the model can be accessed directly by Monte Carlo simulation.

Let us also add that the Dirac operator ${\cal D}$ admits an approximate chirality operator and hence there is an approximate chiral symmetry in this model beside exact rotational invariance,  exact gauge invariance and  exact charge conjugation. The existence of chiral symmetry (though approximate) means that there should exist more structure in the low energy fermionic spectrum.

The partition function $Z$ is also invariant under the translation $X_{\mu}\longrightarrow X_{\mu}+\epsilon X_{\mu}$ where $\epsilon$ is a small parameter.  Under this coordinate transformation the  measure $dX_{\mu}$ changes to $(1+4(N^2-1)\epsilon)dX_{\mu}$. The bosonic action $S_B=S_4+S_3+S_2$ changes to $S_B+\epsilon(4S_4+3S_3+2S_2)$ under $X_{\mu}\longrightarrow X_{\mu}+\epsilon X_{\mu}$. The determinant, on the other hand, changes under $X_{\mu}\longrightarrow X_{\mu}+\epsilon X_{\mu}$ as 
\begin{eqnarray}
{\rm det}{\cal D}
&\longrightarrow & (1+\epsilon)^{2(N^2-1)}{\rm det}^{'}\bigg(1-\frac{\epsilon}{{\cal M}^{'}}\tilde{\xi}\alpha(1+\gamma)\bigg){\rm det}^{'}{\cal M}^{'}.
\end{eqnarray}
The matrices ${\cal M}^{'}$ and $\gamma$ are given in appendix $B$. We obtain then
\begin{eqnarray}
{\rm det}{\cal D}\longrightarrow \bigg(1+\epsilon\bigg[2(N^2-1)-\tilde{\xi}\alpha Tr^{'}_{\rm ad}\frac{1}{\cal D}-\tilde{\xi}\alpha Tr^{'}_{\rm ad}\frac{1}{\cal D}\gamma\bigg]\bigg) {\rm det}{\cal D}.
\end{eqnarray}
From the invariance of the partition function under the coordinate transformation $X_{\mu}\longrightarrow X_{\mu}+\epsilon X_{\mu}$ we derive therefore the Schwinger-Dyson identity

\begin{eqnarray}
{\rm IDE}&=&4\frac{<{\rm YM}>}{N^2}+4\frac{<{\rm YM}_0>}{N^2}+3\frac{<{\rm CS}>}{N^2}+2\frac{<{\rm RAD}>}{N^2}+\tilde{\xi}\alpha \frac{{\rm COND}}{N^2}+\frac{6}{N^2}\nonumber\\
&\equiv &6.\label{ide}
\end{eqnarray}
This is an exact result. 

The operators ${\rm YM}$, ${\rm YM}_0$ and ${\rm CS}$ are the actions given by
\begin{eqnarray}
&&{\rm YM}= -\frac{N}{4}Tr[X_a,X_b]^2~,~{\rm YM}_0= -\frac{N}{2}Tr[X_0,X_a]^2~,~{\rm CS}=\frac{iN\alpha}{3}{\epsilon}_{abc}Tr[X_a,X_b]X_c.\nonumber\\
\end{eqnarray}
The action ${\rm RAD}$ is related to the radius of the sphere. It is given by
\begin{eqnarray}
{\rm RAD}=N\tilde{\beta}\alpha^2  TrX_a^2.
\end{eqnarray}
We will define the radius $r$ of the sphere through the relation
\begin{eqnarray}
\frac{1}{r}=\frac{1}{N\alpha^2c_2}TrX_a^2.
\end{eqnarray}
The total bosonic action is given by $S={\rm YM}+{\rm YM}_0+{\rm CS}+{\rm RAD}$. 
Convergence of Yang-Mills path inetgrals such as the one given by (\ref{pathI}) was studied extensively in \cite{Austing:2001ib}  and in \cite{Austing:2001bd,Krauth:1998xh,Krauth:1999qw, Krauth:1998yu, Austing:2001pk,Austing:2003kd}. This question is of paramount importance for analytical analysis as well as for Monte Carlo simulation. The source of the divergence, if any lies in the so-called flat directions, i.e. the set of commuting matrices. 

The path integral (\ref{pathI}) corresponds to a gauge theory with gauge group $SU(N)$ in dimension $D=4$. We will also consider $SU(N)$ gauge theory in dimension $D=3$.  We start the discussion with the model $\alpha=0$, $\tilde{\beta}=0$ and $\tilde{\xi}=0$. It was found in \cite{Austing:2001bd} that the bosonic path integral in $D=3$ is convergent for $N\geq 4$ while the bosonic path integral in $D=4$ is convergent for $N\geq 3$. Since we are interested in large values of $N$ we can safely consider the bosonic path integrals in $D=3,4$ to be convergent for all practical purposes. On the other hand, it is found in \cite{Austing:2001pk}, that the supersymmetric path integral in $D=3$ is not convergent  while the supersymmetric path integral in $D=4$ is convergent for all $N\geq 2$. 

Tuning the parameters  $\alpha$, $\tilde{\beta}$ does not change this picture. For example it was shown in \cite{Austing:2003kd} that adding a Myers term, i.e. considering a non-zero value of $\alpha$, does not change the convergence properties of the $D-$dimensional Yang-Mills matrix path integral. The point is that the Chern-Simons (Myers) term is always small compared to the quartic Yang-Mills term. The same argument should then lead to the conclusion that adding a bosonic mass term, i.e. if we consider a non-zero $\tilde{\beta}$, will not change the above picture.

Tuning the fermion mass term, i.e. considering a non-zero value of the scalar curvature $\tilde{\xi}$, will lead to complications. In this case the Pfaffian, or equivalently the determinant, will be expanded as a polynomial in the scalar curvature $\tilde{\xi}$. The analysis of \cite{Austing:2001pk} should then be repeated for every term in this expansion. We claim that the supersymmetric path integral in $D=4$ is not convergent  for generic values of $\tilde{\xi}$.

We have extensively checked in Monte Carlo simulation the conjecture that Yang-Mills matrix models in dimension $D=4$ does not make sense  for generic values of $\tilde{\xi}$. The major observation is that for $\tilde{\xi}\neq 0$ the fermion determinant for generic values of $\tilde{\alpha}=\sqrt{N}\alpha$  never reaches thermalization \footnote{This happens typically for small values of $\tilde{\alpha}$ far from the fuzzy sphere region but not very close to $\tilde{\alpha}=0$.}. However, we have also observed that for sufficiently small values  of $\tilde{\xi}$ the theory actually makes sense and thus there is some critical value of $\tilde{\xi}$, which we will not determine in this article, above which the path integral is ill defined. The value of interest $\tilde{\xi}=2/3$ corresponding to the mass deformed matrix model lies in this range where the model is actually ill defined.

Therefore in order to access the mass deformed Yang-Mills matrix model by the Monte Carlo method we must regularize the theory in such a way as to make sure that the path integral is absolutely convergent. Unfortunately most regularizations will not maintain neither the full ${\cal N}=1$ mass deformed supersymmetry nor the half ${\cal N}=1$ cohomologically deformed supersymmetry of this model. We adopt here the regularization in which we will simply set $\tilde{\xi}=0$. In other words we make the replacement

\begin{eqnarray}
S_{\rm SUSY}\longrightarrow S_{\rm SUSY}^{'}&=&NTr\bigg[-\frac{1}{4}[X_a,X_b]^2+\frac{2i\alpha}{3}{\epsilon}_{abc}X_aX_bX_c\bigg]+N\tilde{\beta}\alpha^2TrX_a^2\nonumber\\
&-&\frac{1}{N\alpha}Tr{\theta}^+\bigg(i[X_4,..]+{\sigma}_a[X_a,..]\bigg).\label{action1}
\end{eqnarray}
In summary, the value $\tilde{\beta}=2/9$ corresponds to the mass deformed Yang-Mills matrix model with softly broken supersymmetry whereas the value $\tilde{\beta}=0$ is precisely the minimally deformed model which enjoys  half of the ${\cal N}=1$ cohomologically deformed supersymmetry.


\subsection{Bosonic Theory: Emergent Geometry and Phase Diagram}
\paragraph{Emergent Geometry}
We measure the different observables as a function of the coupling constant $\tilde{\alpha}$ for the two relevant values of $\tilde{\beta}$, i.e. $\tilde{\beta}=0,2/9$. The first observable is the bosonic Schwinger-Dyson equation given by

\begin{eqnarray}
{\rm IDE}&=&4\frac{<{\rm YM}>}{N^2}+4\frac{<{\rm YM}_0>}{N^2}+3\frac{<{\rm CS}>}{N^2}+2\frac{<{\rm RAD}>}{N^2}+\frac{4}{N^2}
\equiv 4.\label{ideB}
\end{eqnarray}
We have verified that the bosonic Schwinger-Dyson equation holds in Monte Carlo simulations to a very satisfactory accuracy. 

The radius which we have defined by the equation $<1/r>=<TrX_a^2>/\tilde{\alpha}^2c_2$ is shown on figures (\ref{radiusF1}) and (\ref{radiusF2}). 
For large values of $\tilde{\alpha}$ the result is consistent with the classical prediction
\begin{eqnarray}
<\frac{1}{r}>=<\frac{TrX_a^2}{\tilde{\alpha}^2c_2}>=\phi_+^2~,~\phi_+=\frac{1+\sqrt{1-4\tilde{\beta}}}{2}.
\end{eqnarray}
This means in particular that the system is in the ground state configurations
\begin{eqnarray}
X_4=0~,~X_a=\alpha\phi L_a.
\end{eqnarray}
In other words we have a fuzzy spherical geometry given by the commutation relations
\begin{eqnarray}
[X_4,X_a]=0~,~[X_a,X_b]=i\epsilon_{abc}\alpha\phi X_c.
\end{eqnarray}
We have checked these commutation relations and found them to hold quite well for sufficiently large values of $\tilde{\alpha}$. The coordinates on the sphere are defined by
\begin{eqnarray}
n_a=\frac{X_a}{\sqrt{c_2}\alpha}~,~\sum_an_a^2=\phi^2.
\end{eqnarray}
We observe that as we decrease $\tilde{\alpha}$, the radius $1/r$ jumps abruptly to $0$ then starts to increase again until it becomes infinite at $\tilde{\alpha}=0$. This is the most dramatic effect of the so-called sphere-to-matrix transition in which the sphere suddenly expands  and evaporates at the transition points then it starts shrinking to zero rapidly as we lower the coupling further.  

This is the interpretation advocated in \cite{DelgadilloBlando:2008vi,DelgadilloBlando:2007vx,O'Connor:2006wv} for a similar phenomena observed in the case of three dimensional bosonic models. As far as we know this phenomena was observed in Monte Carlo simulation first in \cite{Azuma:2004zq} and it was found in analytical work on perturbative three dimensional bosonic models in \cite{CastroVillarreal:2004vh} and then in \cite{Azuma:2004ie}.

The transitions for the bosonic mass deformed model with $\tilde{\beta}=2/9$ and the bosonic cohomological model with $\tilde{\beta}=0$ are observed to occur at the following estimated values

\begin{eqnarray}
\tilde{\alpha}_*=4.9\pm 0.1~,~\tilde{\beta}=2/9.
\end{eqnarray}
  \begin{eqnarray}
\tilde{\alpha}_*=2.55\pm 0.1~,~\tilde{\beta}=0.
\end{eqnarray}
The fuzzy sphere phase corresponds to the region $\tilde{\alpha}>\tilde{\alpha}_*$ whereas the matrix phase corresponds to the region  $\tilde{\alpha}<\tilde{\alpha}_*$. In other words the sphere becomes more stable as we make $\tilde{\beta}$ smaller (see below). 

The order of the sphere-to-matrix transition is very difficult to determine. Since the ground state configurations are $X_4=0$ and $X_a=\alpha\phi L_a$, the theoretical analysis based on the effective potential of the three dimensional model done in \cite{DelgadilloBlando:2008vi,DelgadilloBlando:2007vx} should also hold here largely unchanged (see below). As a consequence we will only summarize here the main points omitting much technical details.

\begin{figure}[htbp]
\begin{center}
\includegraphics[width=10.0cm,angle=-90]{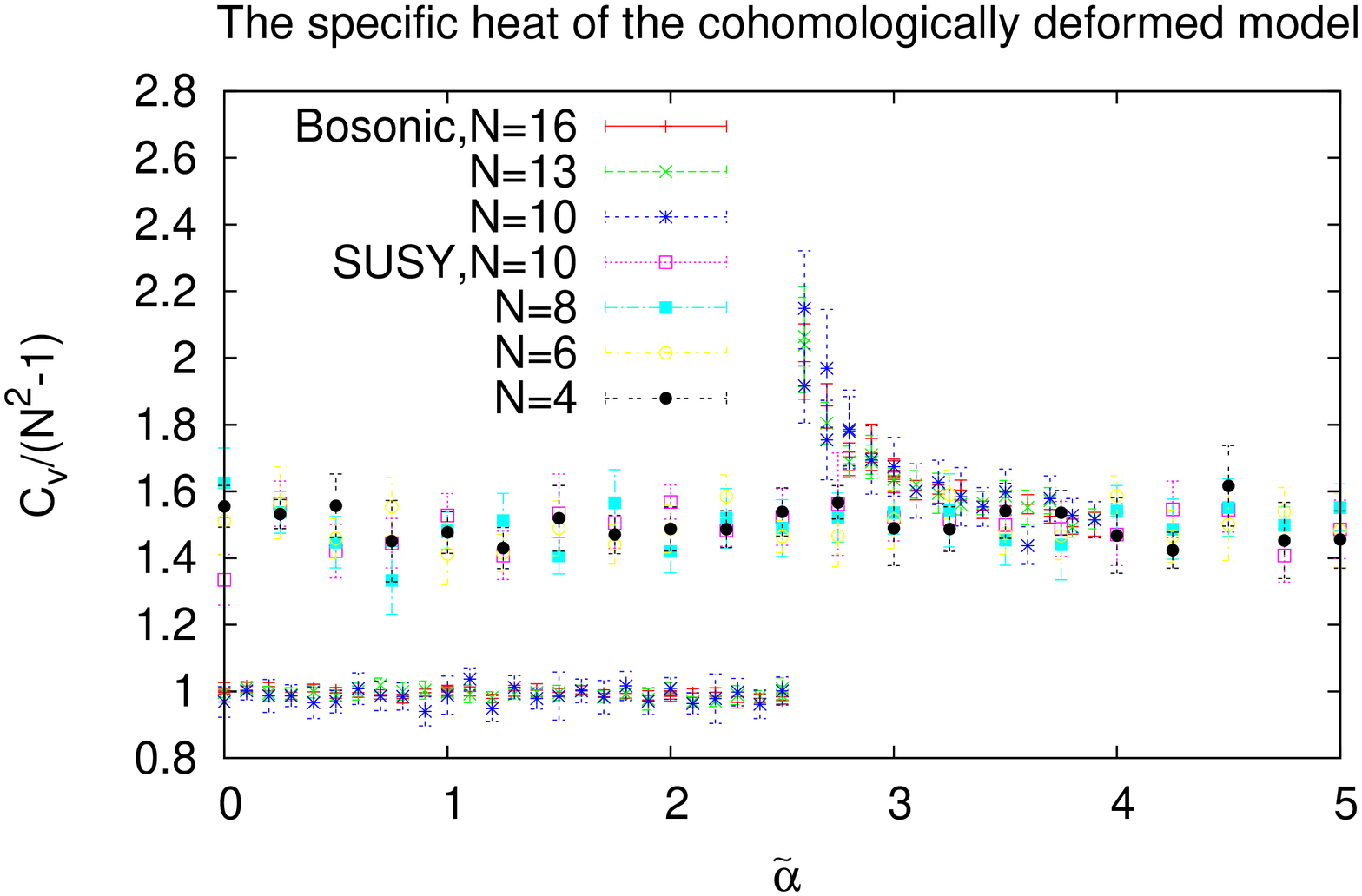}
\includegraphics[width=10.0cm,angle=-90]{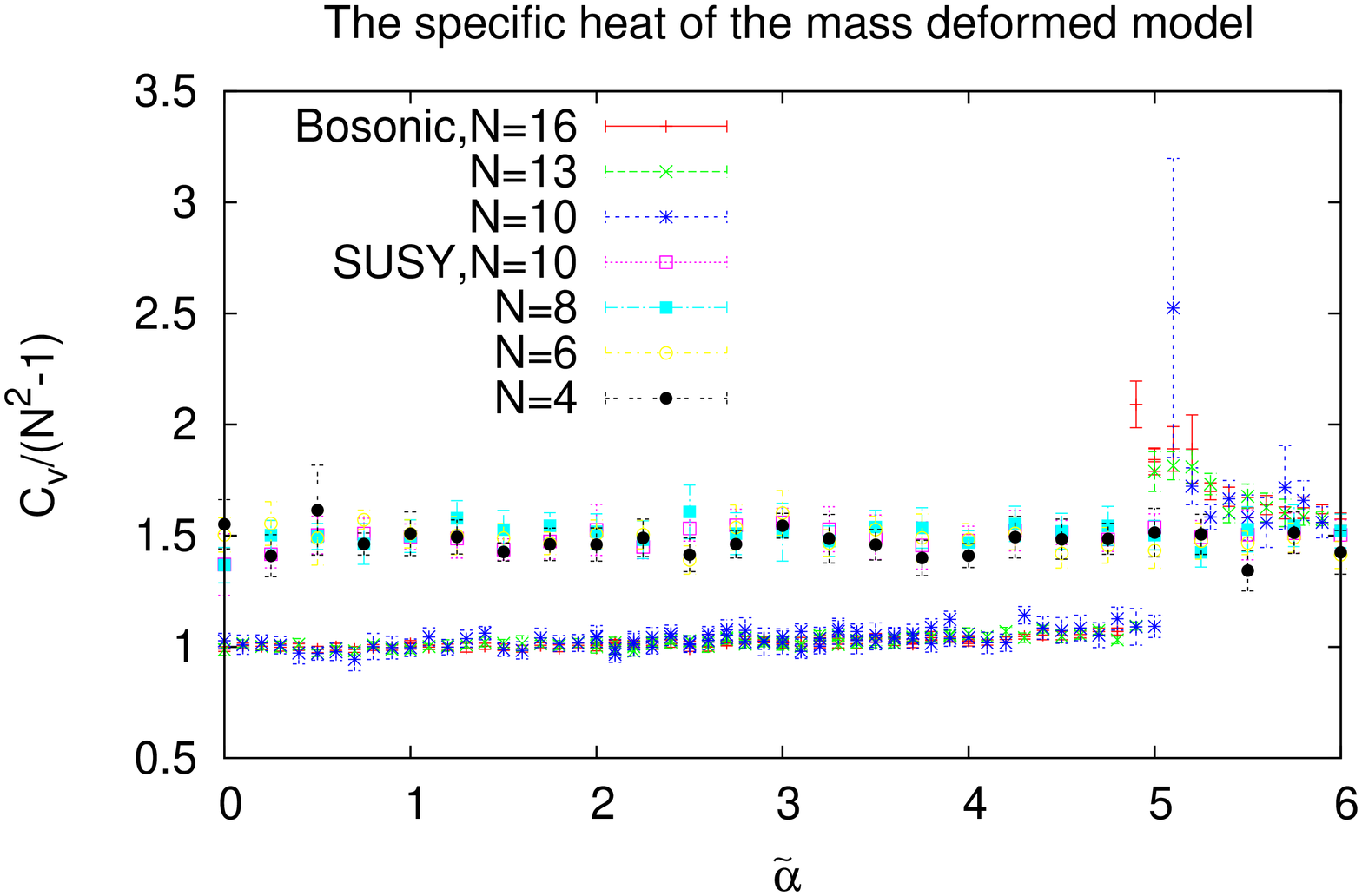}
\caption{The specific heat of the supersymmetric models. }\label{cvF}
\end{center}
\end{figure}

The specific heat $C_v=(<S^2>-<S>^2)/(N^2-1)$ shown on figure (\ref{cvF}) diverges from the side of the fuzzy sphere with a critical exponent equal $1/2$. It is equal to $3$ in this phase where $1/2$ is due to the $2$ dimensional $U(1)$ gauge field on the sphere, $1/2$ is due to the normal scalar field on the sphere and $1/2$ is due to the scalar field $X_4$.  This critical behavior is typical of a second order transition. In the matrix phase the specific heat is a constant right up to the transition point and it is equal $1$ where each matrix contributes $1/4$. There is no divergence from this side and the critical exponent is $0$. In other words the behavior above and below the critical coupling are different, which is quite unusual, but still from the specific heat we qualify this transition as second order.  

\begin{figure}[htbp]
\begin{center}
\includegraphics[width=10.0cm,angle=-90]{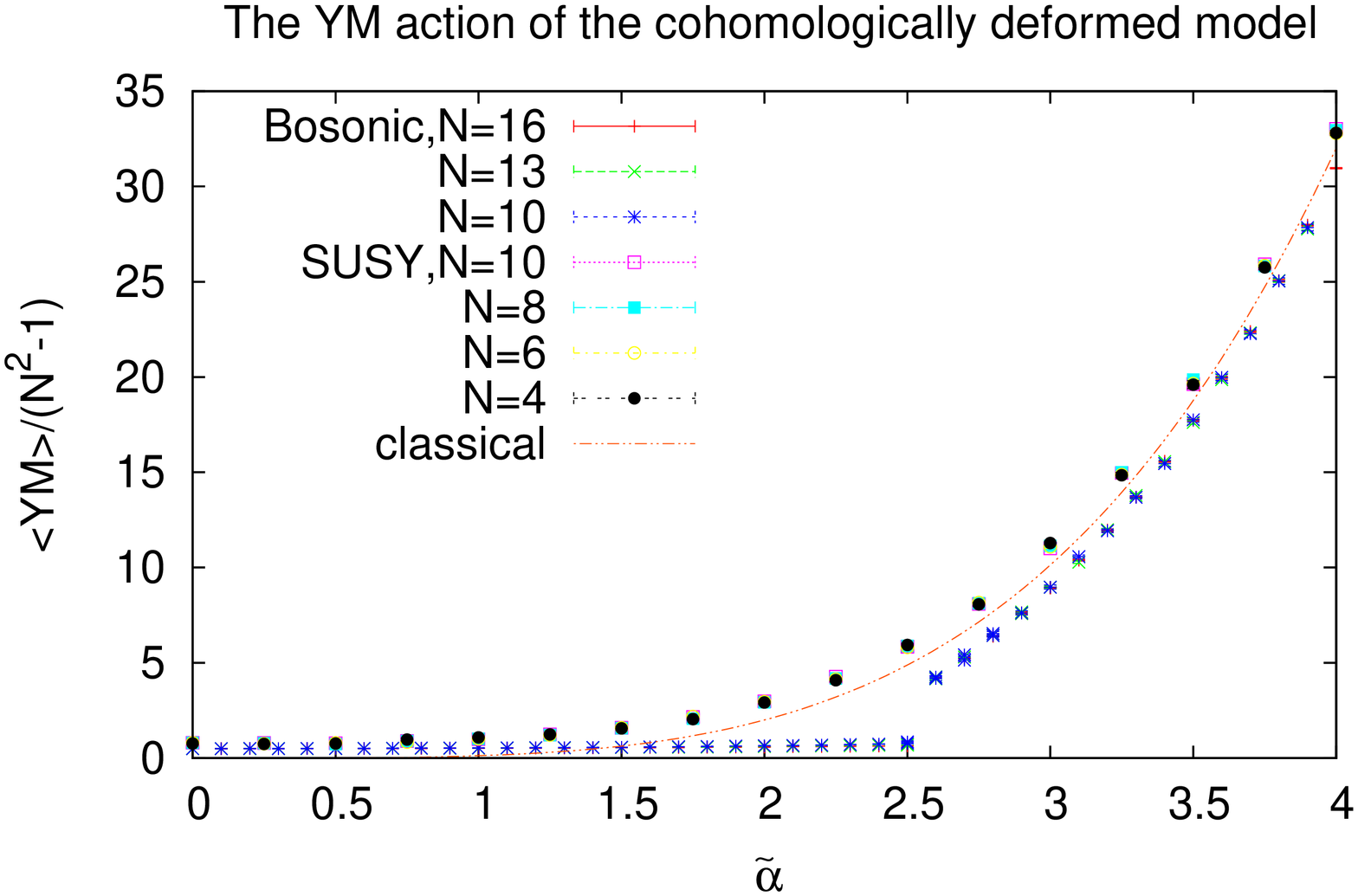}
\includegraphics[width=10.0cm,angle=-90]{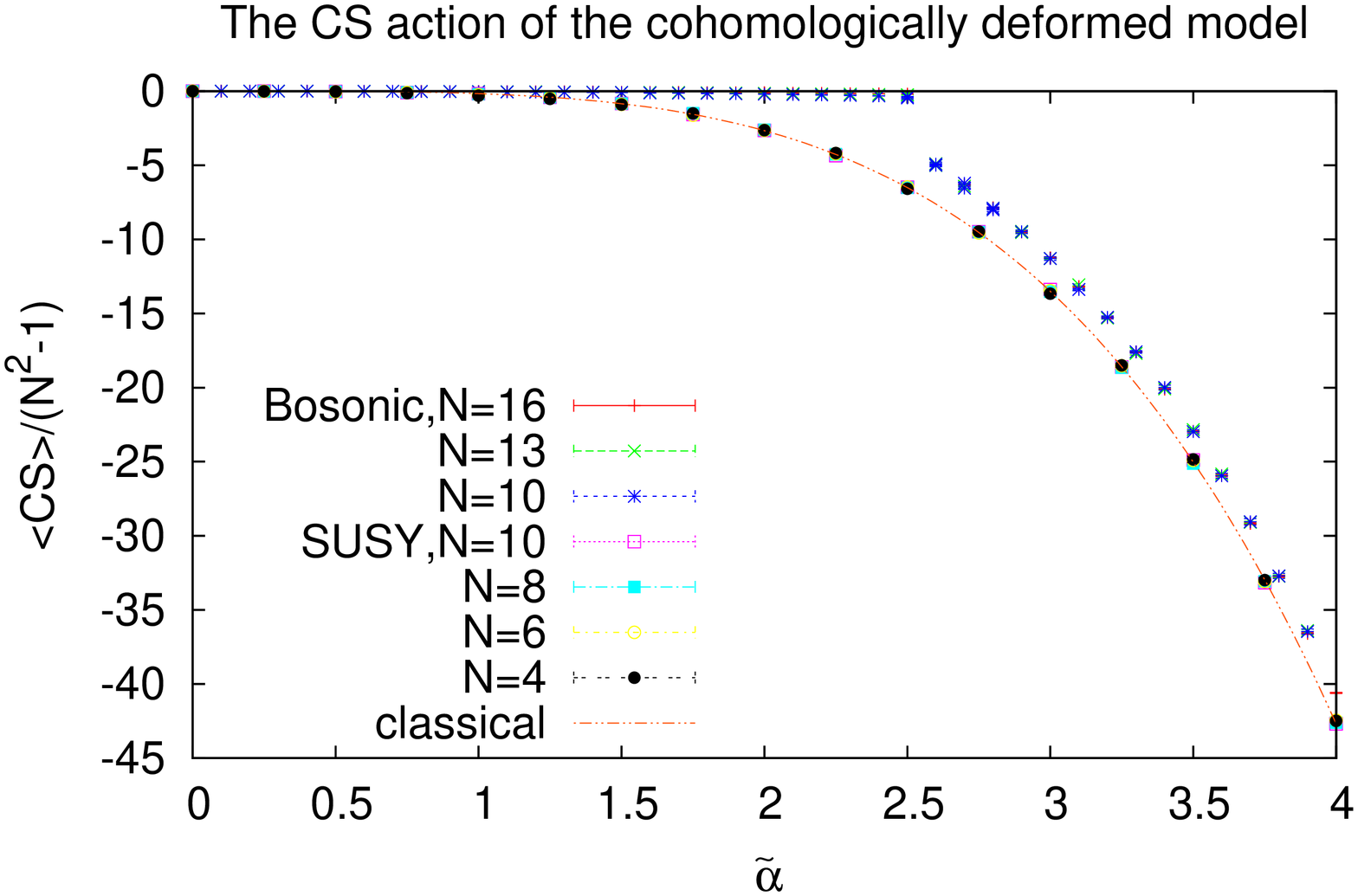}
\caption{The average of the Yang-Mills and Myers actions of the cohomologically deformed  model. }\label{YMCSF1}
\end{center}
\end{figure}

\begin{figure}[htbp]
\begin{center}
\includegraphics[width=10.0cm,angle=-90]{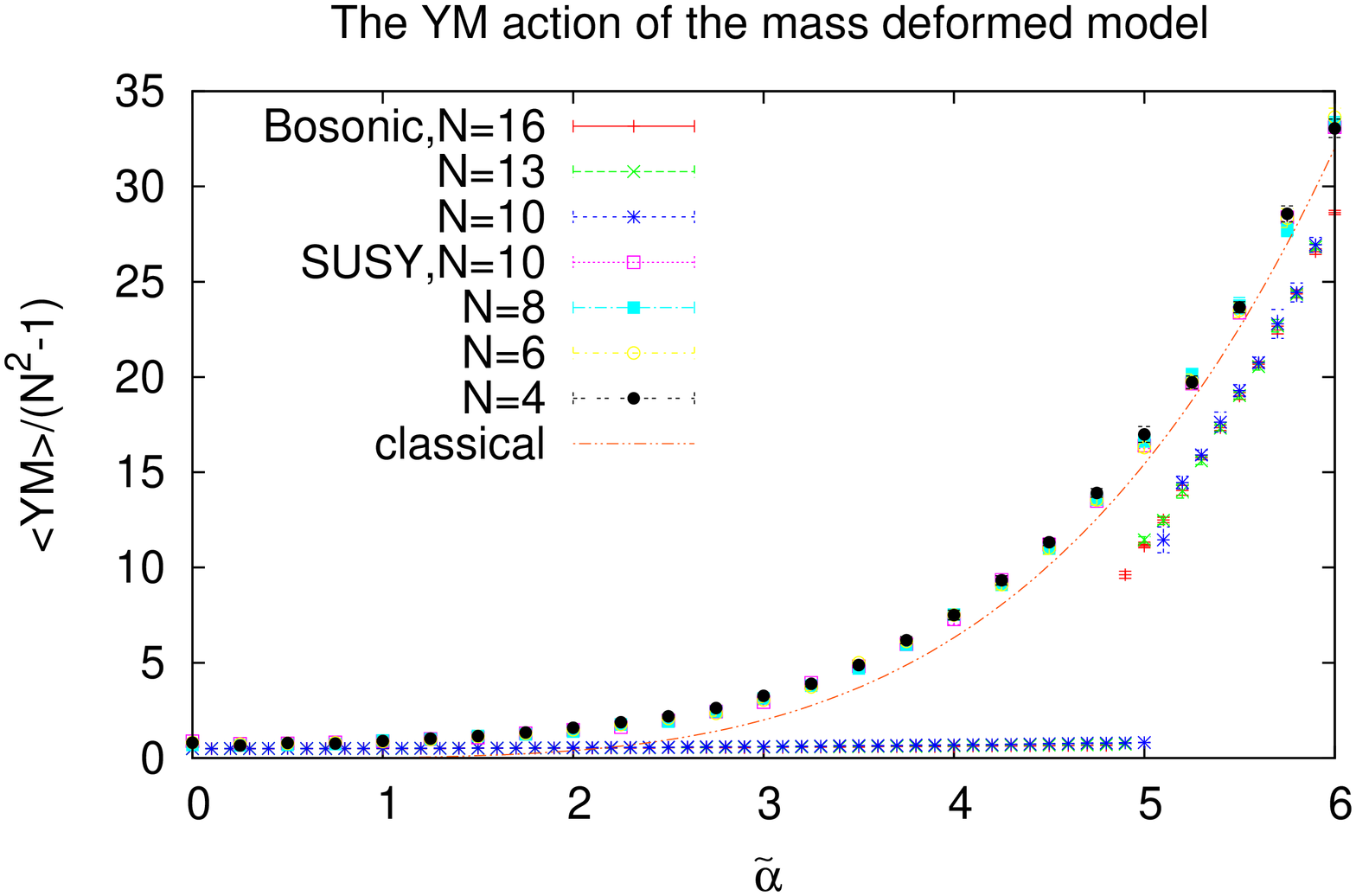}
\includegraphics[width=10.0cm,angle=-90]{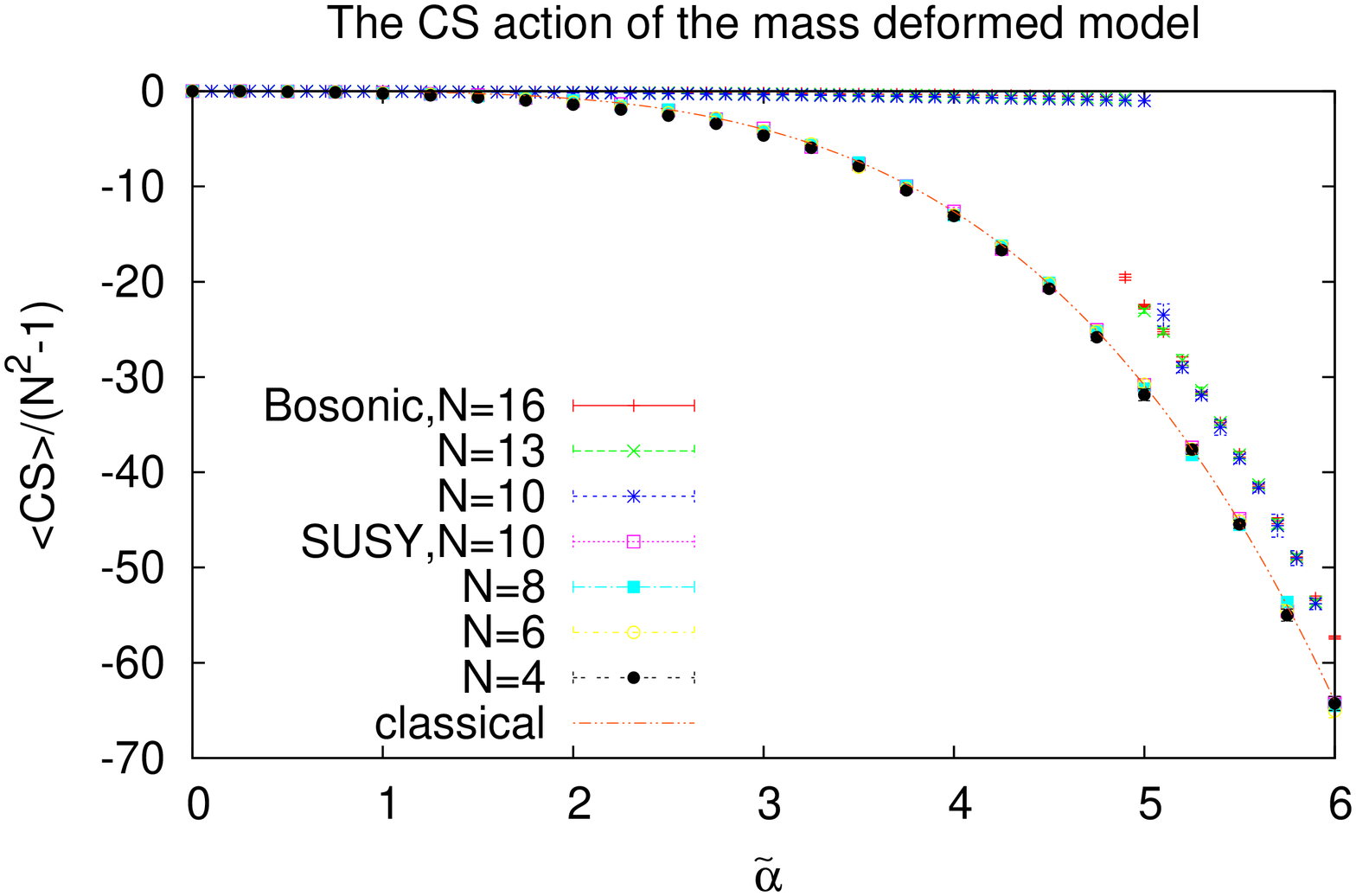}
\caption{The average of the Yang-Mills and Myers actions of the  mass deformed model. }\label{YMCSF2}
\end{center}
\end{figure}

\begin{figure}[htbp]
\begin{center}
\includegraphics[width=10.0cm,angle=-90]{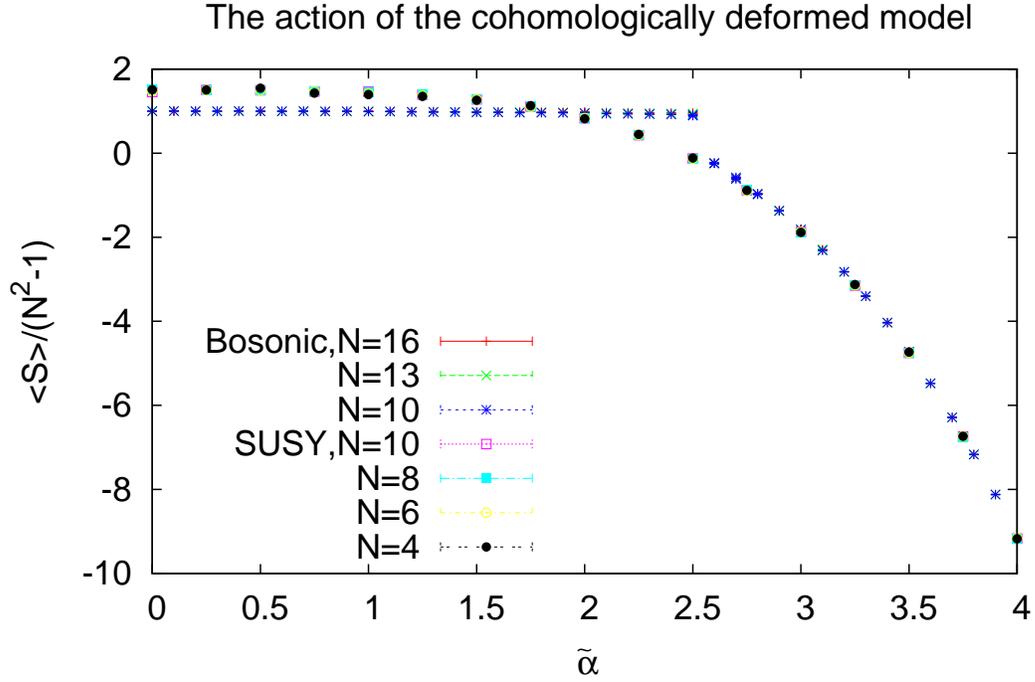}
\includegraphics[width=10.0cm,angle=-90]{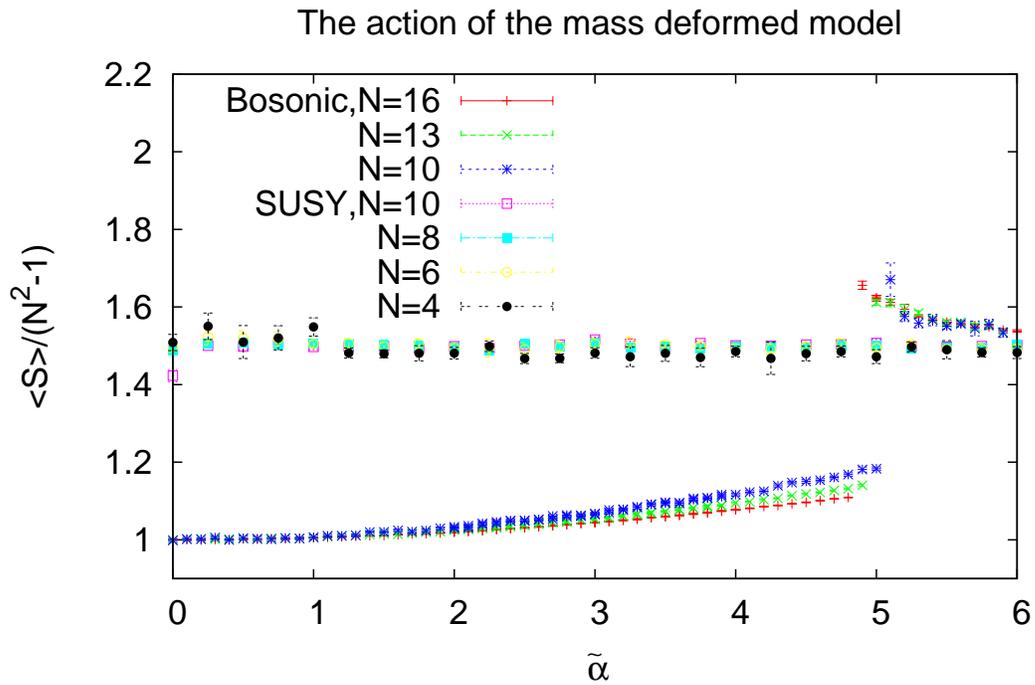}
\caption{The average of the total action of the supersymmetric models. }\label{actionF}
\end{center}
\end{figure}

The expectation values of the Yang-Mills action and the Myers (or Chern-Simons) action are shown on (\ref{YMCSF1}) and (\ref{YMCSF2}). The expectation values of the total bosonic action for the two cases  $\tilde{\beta}=2/9$ and $\tilde{\beta}=0$ are shown on figure (\ref{actionF}). From these observables we observe a discontinuity at the transition point. Thus the sphere-to-matrix transition is associated with  a latent heat equal to $\Delta <S>=<S>_{\rm matrix}-<S>_{\rm sphere}$ which is typical of a first order phase transition. It is straightforward to estimate the value of this latent heat. The latent heat is released by going from the matrix phase to the fuzzy sphere phase  for $\tilde{\beta}=0$ whereas for $\tilde{\beta}=2/9$ the latent heat is released by going in the other direction from the fuzzy sphere phase to the matrix phase.

As we will see from the discussion of the eigenvalues distributions  the matrices $X_{\mu}$ in the matrix phase are commuting matrices centered around $0$.
\paragraph{Phase Diagram}
The last point we would like to address  within the context of the bosonic theory is the construction of the phase diagram in the plane $\tilde{\alpha}-\tilde{\beta}$. We have already measured two points of this phase diagram which correspond to the two values  $\tilde{\beta}=2/9$ and $\tilde{\beta}=0$.  In order to map the phase boundary between the fuzzy sphere and the matrix phase we choose other values of $\tilde{\beta}$ and  measure for each one of them the critical value of $\tilde{\alpha}$ from the discontinuity in the radius $1/r$. 

The effective potential of the $4$ dimensional bosonic Yang-Mills matrix model in the Feynman-'t Hooft background field gauge in the  ground state configurations $X_4=0$ and $X_a=\alpha\phi L_a$ can be calculated using the method of \cite{CastroVillarreal:2004vh}. We find
  \begin{eqnarray}
\frac{V_{\rm eff}}{2c_2}=\tilde{\alpha}^4\bigg[\frac{\phi^4}{4}-\frac{\phi^3}{3}+\tilde{\beta}\frac{\phi^2}{2}\bigg]+2\log\phi^2.
\end{eqnarray}
The difference with the three dimensional bosonic Yang-Mills matrix model lies in the factor of $2$ multiplying the logarithm. The critical line can then be obtained following the method of  \cite{DelgadilloBlando:2008vi}. We get
 \begin{eqnarray}
\phi_*=\frac{3}{8}(1+\sqrt{1-\frac{32\tilde{\beta}}{9}})~,~\tilde{\alpha}_*^4=\frac{16}{\phi_*^2(\phi_*^2-2\tilde{\beta})}.
\end{eqnarray}
This prediction is in a  very reasonable agreement with the Monte Carlo data.
\subsection{Dynamical Fermions: Impact of Supersymmetry}

In this section we will discuss the effect of the fermion determinant. First we note that simulations with fermions are much more harder than pure bosonic simulations. The main source of complication is the evaluation of the determinant which is highly non trivial. Thermalization is very difficult and as a consequence taking the limit of large $N$ is not so easy even with the use of the Hybrid Monte Carlo algorithm. In the bosonic case we could go as large as $N=100$ with very decent number of statistics although in this article we have only reported data with $N$ up to $N=16$. In the fermionic case we will report data with $N$ up to $N=10$. 

The first thing we have checked is the Schwinger-Dyson identity (\ref{ide}). The supersymmetric Monte Carlo data agrees well with the prediction $6$ as opposed to the bosonic data which agrees with the prediction $4$. Note that $6=4+2$ where $4$ is the number of bosonic matrices and $2$ is the number of fermionic matrices.

\begin{figure}[htbp]
\begin{center}
\includegraphics[width=10.0cm,angle=-90]{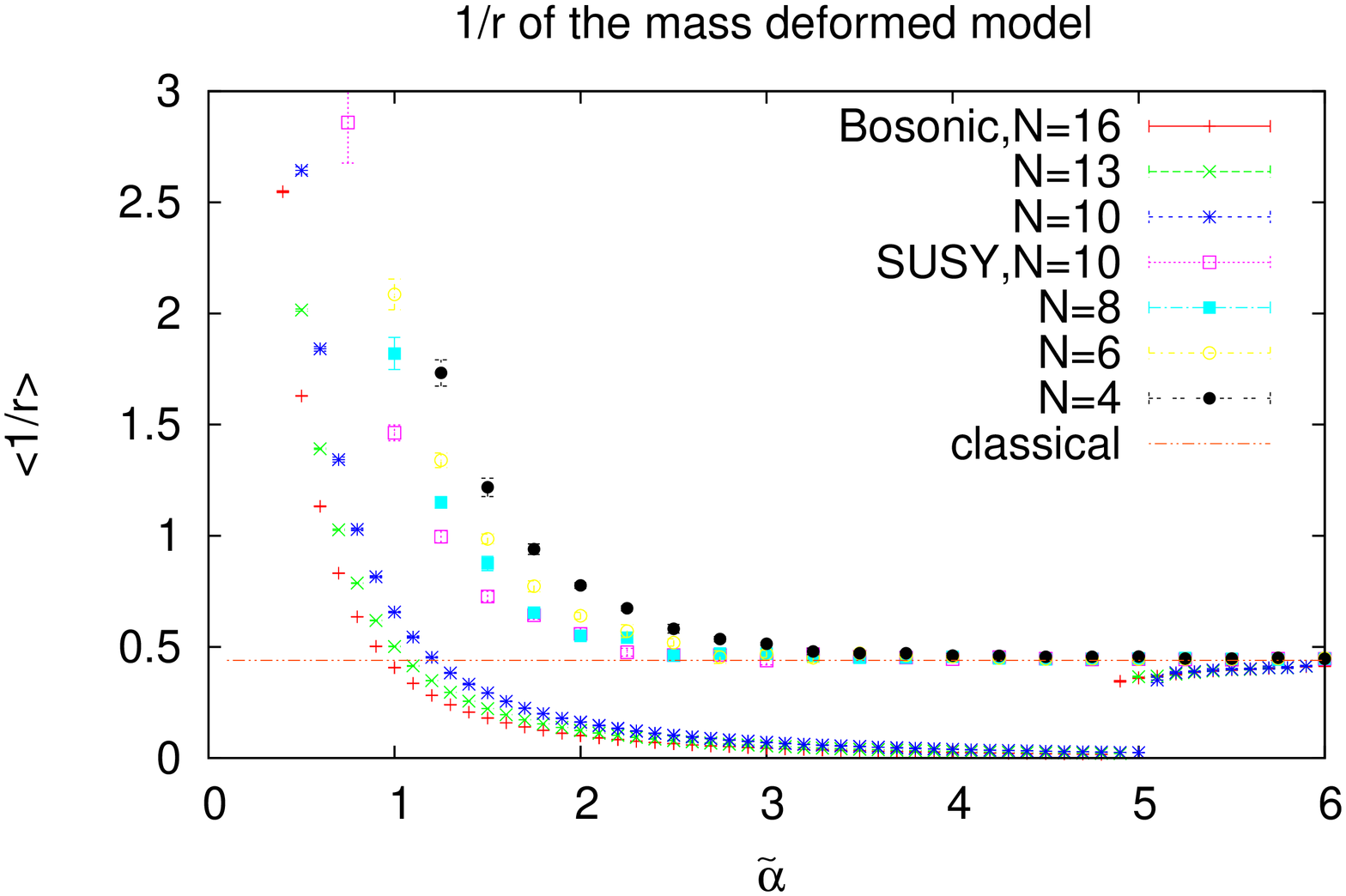}
\includegraphics[width=10.0cm,angle=-90]{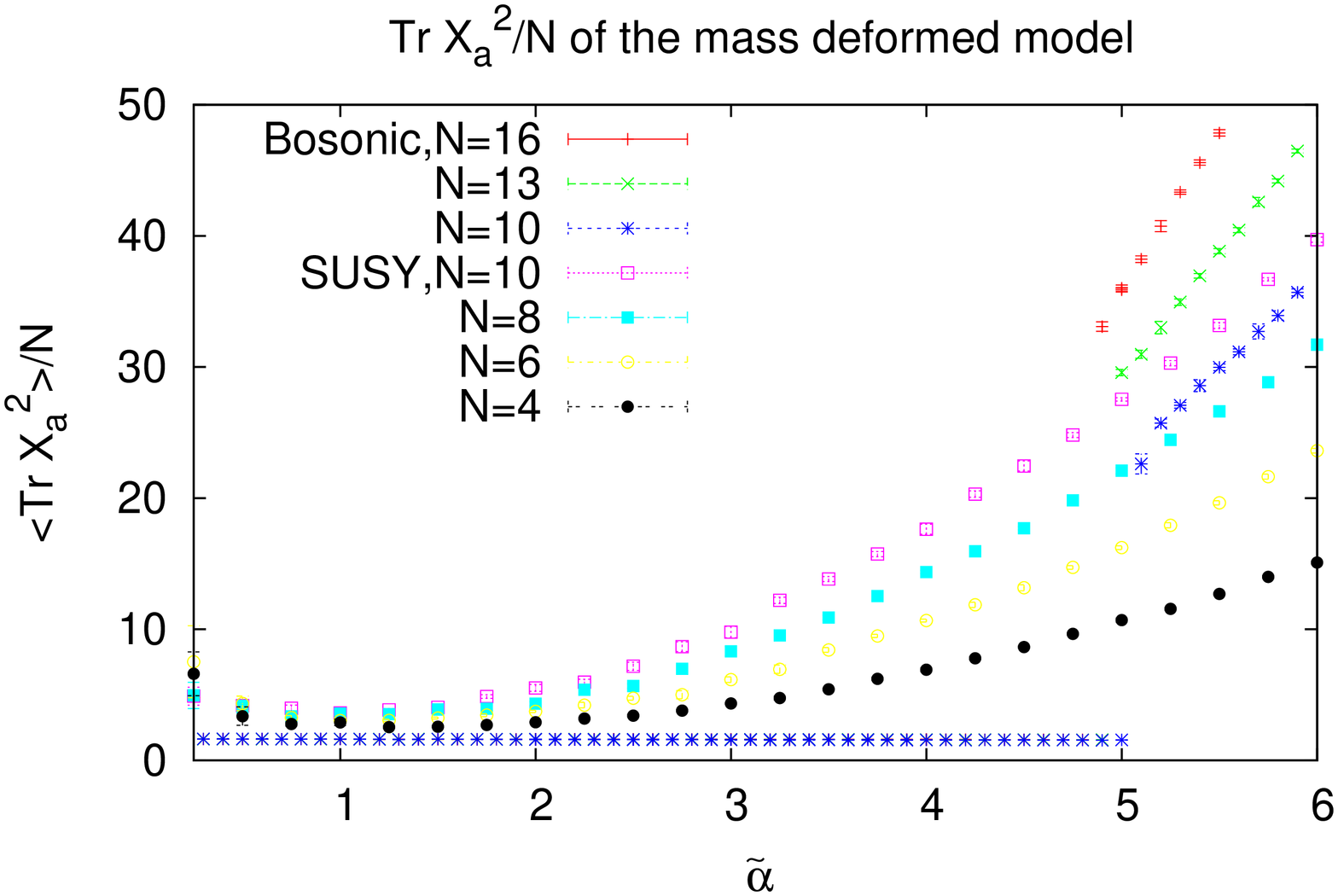}
\caption{The radius of the mass deformed model. }\label{radiusF1}
\end{center}
\end{figure}

\begin{figure}[htbp]
\begin{center}
\includegraphics[width=10.0cm,angle=-90]{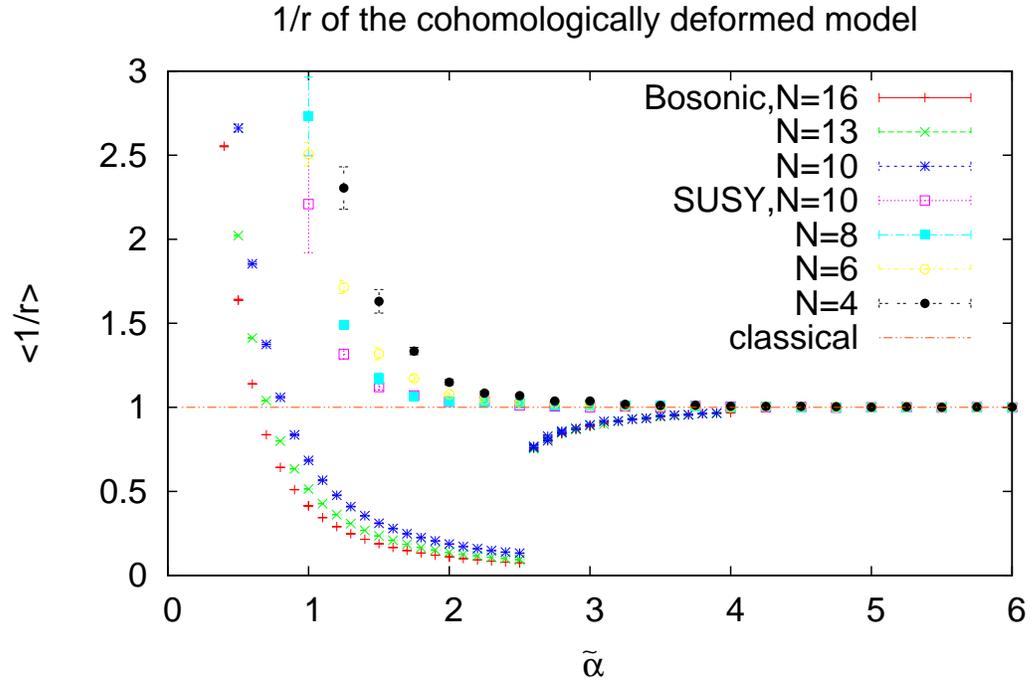}
\includegraphics[width=10.0cm,angle=-90]{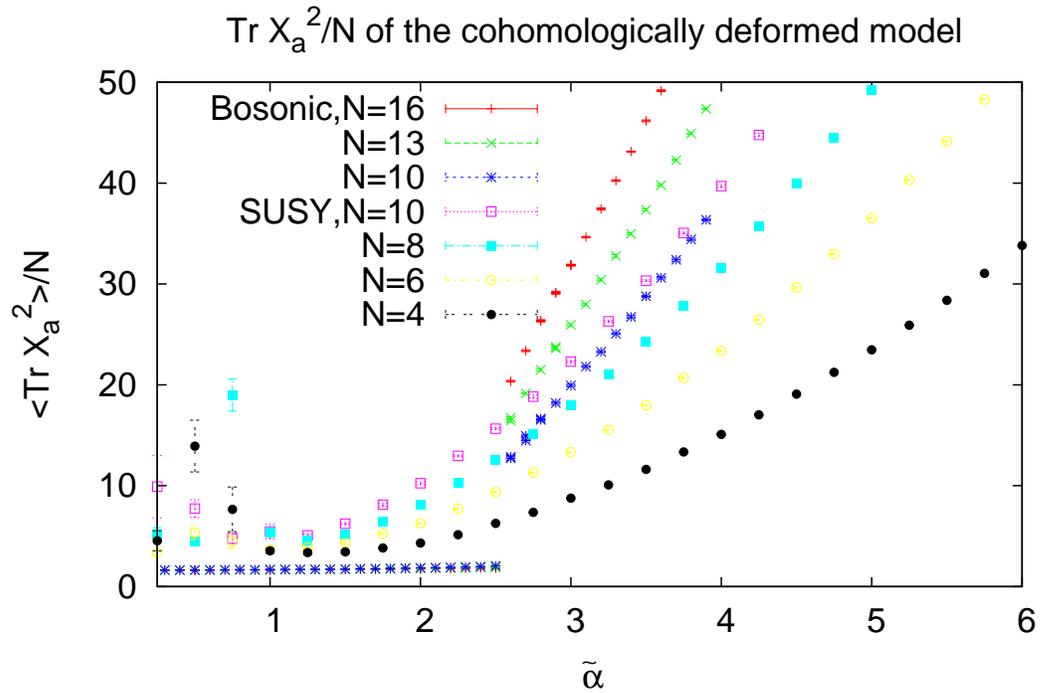}
\caption{The radius of the cohomologically deformed model. }\label{radiusF2}
\end{center}
\end{figure}

The most important order parameter with direct significance to the underlying geometry is the radius $1/r$. See again figures (\ref{radiusF1}) and  (\ref{radiusF2}). We observe that the transition sphere-to-matrix observed in the bosonic theory disappeared completely. Again it seems here that there is no major difference between the two models with $\tilde{\beta}=2/9$ and $\tilde{\beta}=0$. It is clear from the structure of the action that the theory with $\tilde{\alpha}=0$ can not support the fuzzy sphere geometry and thus a phase with commuting matrices must still exist. However, the transition to the phase of commuting matrices starting from the fuzzy sphere phase seems to be a crossover transition not the second/first order behavior observer in the bosonic model. This seems to be confirmed by the behavior of the specific heat, the Yang-Mills and Myers actions and the total action shown on figures  (\ref{cvF}), (\ref{YMCSF1}-\ref{YMCSF2}) and (\ref{actionF}) respectively. The jump and critical behavior in the specific heat and the discontinuity in the various actions disappeared.

We have to note here that the observable $<Tr X_a^2>/N$ diverges in the supersymmetric theory with $\tilde{\alpha}=\tilde{\beta}=0$ \cite{Krauth:1999qw}. For the mass deformed theory we have $\tilde{\beta}=2/9$ and thus the observable $<Tr X_a^2>/N$ always exists. We observe on the second graph of figure (\ref{radiusF1}) that $<Tr X_a^2>/N$ increases as we decrease $\tilde{\alpha}$ towards $0$ which is consistent with the fact that it will diverge in the limit $\tilde{\alpha}\longrightarrow 0$. Qualitatively the same phenomena is observed for $\tilde{\beta}=0$ on the second graph of figure (\ref{radiusF2}) with more erratic behavior as we decrease  $\tilde{\alpha}$ towards $0$. However in this case we can not infer that $<Tr X_a^2>/N$ exists for all $\tilde{\alpha}$ since $\tilde{\beta}=0$ although it looks that it does from the data. From this perspective the mass deformed model is better than the cohomologically deformed model.

We have not succeeded in determining precisely the value at which the crossover transition occurs  but it seems that it depends on $N$ in such a way that it is pushed to smaller values of $\tilde{\alpha}$ as we increase $N$. From figures (\ref{radiusF1}) we can read that the crossover transition for  $\tilde{\beta}=2/9$ occurs at $\tilde{\alpha}=3.13,2.63,2.38$ and $2.13$ for $N=4,6,8$ and $N=10$ respectively. A simple fit yields the result
\begin{eqnarray}
\tilde{\alpha}_*^4=\frac{61.13}{N^{2.38}}~,~\tilde{\beta}=2/9.
\end{eqnarray}
The conjecture that the crossover transition occurs at arbitrarily small values of $\tilde{\alpha}$ in the large $N$ limit is one of the main results of this article. In this way the fuzzy sphere is truly stable in the supersymmetric theory and does not decay. In any case we are certain that the fuzzy sphere in the supersymmetric theory is more stable compared to the bosonic theory and the crossover transition to the matrix phase is much slower. This conclusion is similar to that of \cite{Anagnostopoulos:2005cy}.

\subsection{Eigenvalues Distributions} 
\paragraph{Bosonic Theory}
A powerful set of order parameters is given by the eigenvalues distributions of the two matrices $X_3$ and $X_4$. The eigenvalues distribution of the matrix $X_4$ is qualitatively the same for all values of the coupling constants $\tilde{\alpha}$.  However, the eigenvalues distribution of the matrix $X_3$ suffers a major change as we go across the transition point. In the fuzzy sphere region the eigenvalues distribution of $X_3$ is given by an $N-$cut distribution corresponding to the $N$ eigenvalues $-(N-1)/2,...,(N-1)/2$ whereas in the matrix phase  the  eigenvalues distribution of $X_3$ is identical to the  eigenvalues distribution of $X_4$.

The eigenvalues distribution $\rho_4(x_4)$ of the matrix $X_4$ is always centered around $0$. In the fuzzy sphere phase $\rho_4(x_4)$ depends on the coupling constant  $\tilde{\alpha}$. In the matrix phase below the critical value the eigenvalues distribution $\rho_4(x_4)$ does not depend on $\tilde{\alpha}$ and coincides with the eigenvalues distribution of the non deformed model with  $\tilde{\alpha}=0$. In this region the eigenvalues distributions of the matrices $X_3$ and $X_4$ are identical.

\begin{figure}[htbp]
\begin{center}
\includegraphics[width=10.0cm,angle=-90]{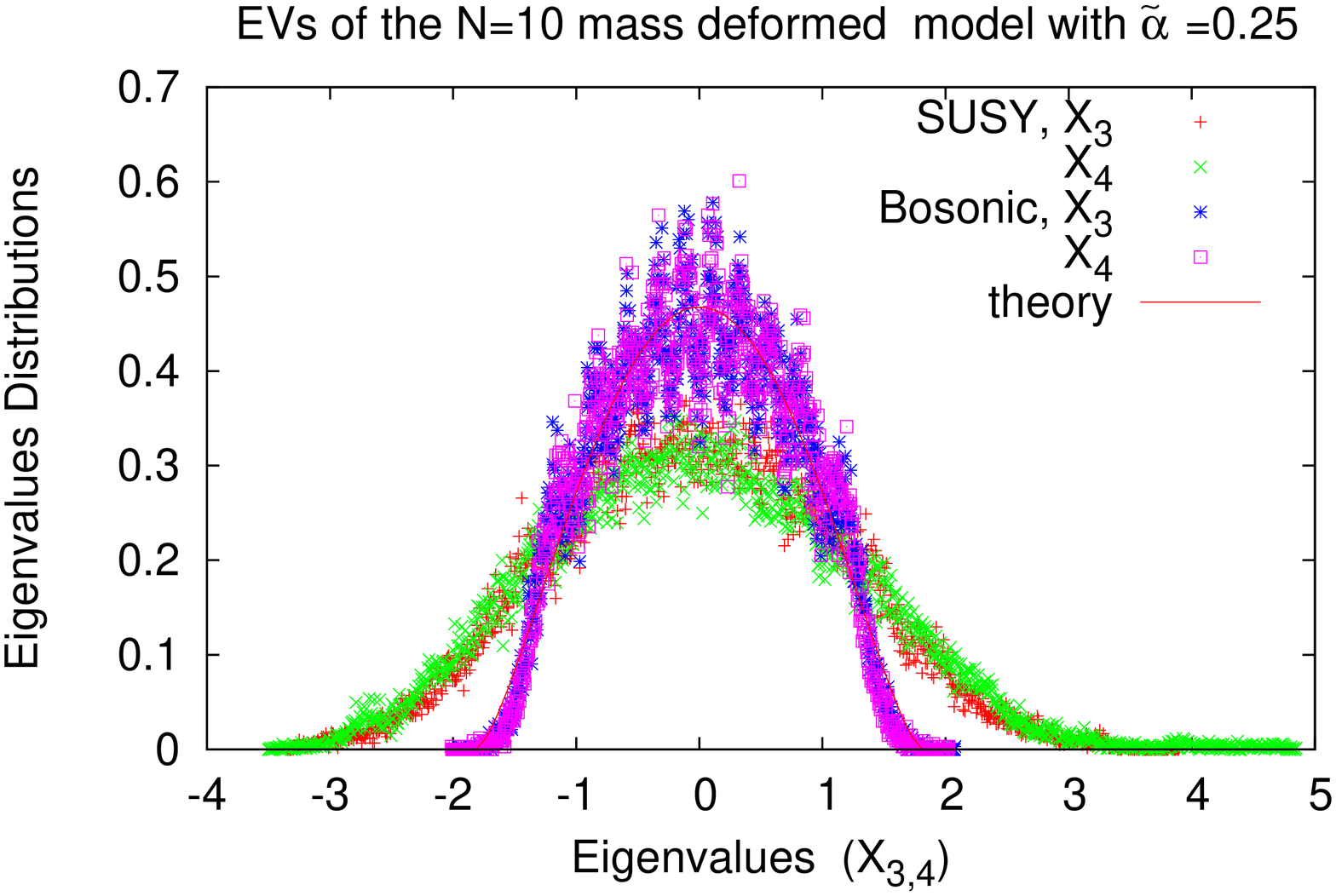}
\includegraphics[width=10.0cm,angle=-90]{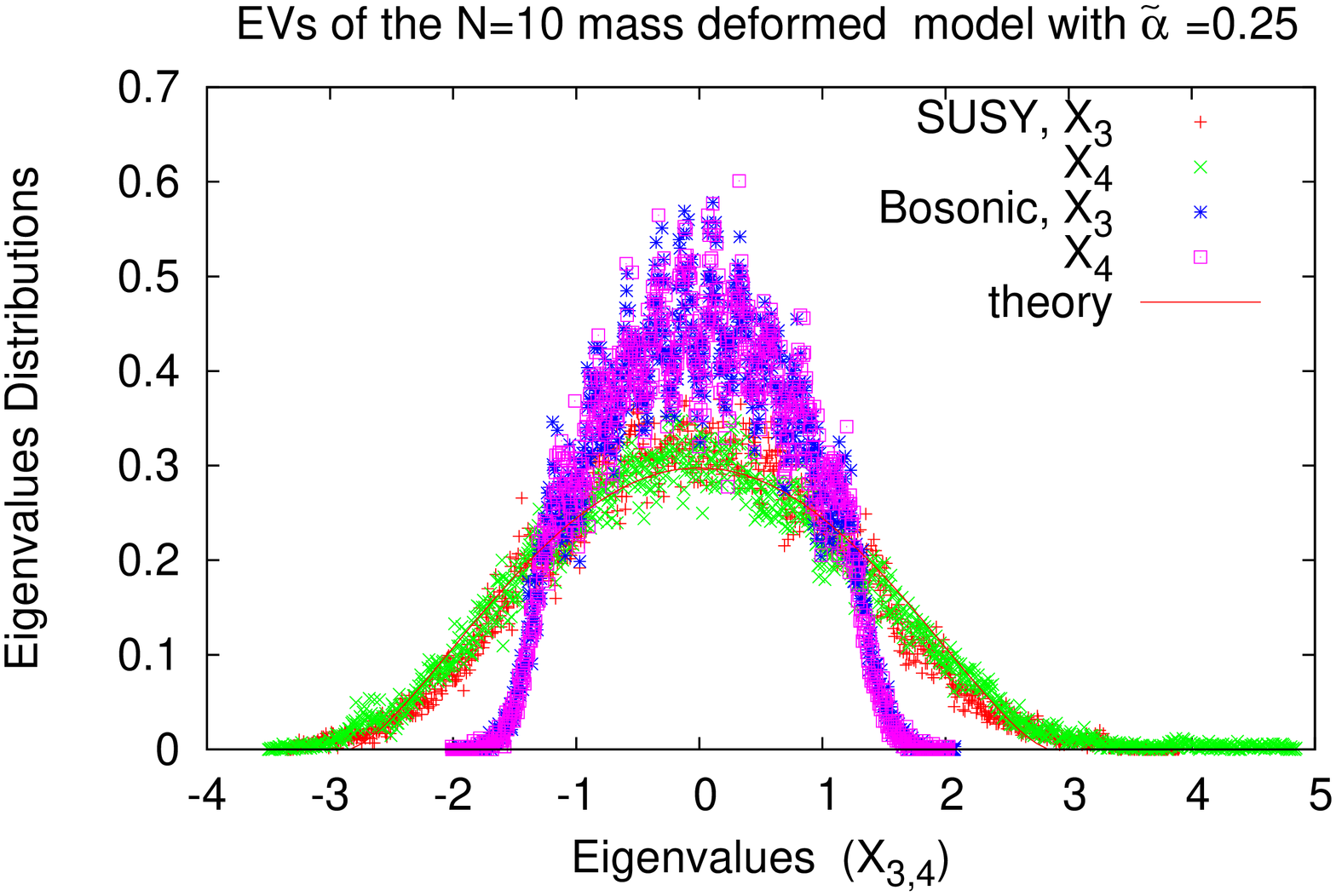}
\caption{The eigenvalues distributions of the mass deformed model. }\label{distrF2}
\end{center}
\end{figure}

 Motivated by the work \cite{Berenstein:2008eg,Hotta:1998en} it was conjectured \cite{denjoeprivate} that the joint eigenvalues distribution of $d$ matrices $X_1$, $X_2$,...$X_d$ with dynamics given by a reduced Yang-Mills action should be uniform inside a solid ball of some radius $R$. We have already checked that this conjecture works in three dimensions \cite{Ydri:2010kg}. We will check now that this conjecture holds also true in four dimensions. Let $\rho(x_1,x_2,x_3,x_4)$ be the joint eigenvalues distribution of the $4$ matrices $X_1$, $X_2$, $X_3$ and $X_4$. We assume that $\rho(x_1,x_2,x_3,x_4)$ is uniform inside a four dimensional ball of radius $R$. The normalization condition gives $\rho(x_1,x_2,x_3,x_4)=1/V_4=2/\pi^2 R^4$. We want to compute the eigenvalues distribution of a single matrix, say $X_4$, which is induced by integrating out the other three matrices. We compute 

 \begin{eqnarray}
\bigg[\int_{-R}^R dx_4\int_{-R}^R dx_3\int_{-R}^R dx_2\int_{-R}^R dx_1\bigg]_{x_1^2+x_2^2+x_3^2+x_4^2\leq R^2} &=& \frac{4\pi}{3}\int_{-R}^R dx_4(R^2-x_4^2)^{\frac{3}{2}}.\nonumber\\
\end{eqnarray}
We obtain therefore the eigenvalues distribution
\begin{eqnarray}
\rho_4(x_4)=\frac{8}{3\pi R^4}(R^2-x_4^2)^{\frac{3}{2}}.\label{4ball}
\end{eqnarray}
This is precisely the fit with a measured value of $R$ for $\tilde{\beta}=0$ and $\tilde{\alpha}=0$ given by
\begin{eqnarray}
R_0=1.826\pm 0.004.
\end{eqnarray}
The above eigenvalues distribution works better for the theory with $\tilde{\beta}=2/9$ as shown on figure (\ref{distrF2}) with a similar measured value of $R$ given for $\tilde{\alpha}=0.25$ by
\begin{eqnarray}
R_{2/9}=1.815\pm 0.008.
\end{eqnarray}
We have found that these two measured values are almost the same throughout the matrix phase.

We emphasize that $\rho_4(x_4)$ is the eigenvalues distribution of the matrix $X_4$ not only in the matrix phase but also in the fuzzy sphere phase with a value of $R$ which depends on the coupling constant $\tilde{\alpha}$. We also emphasize that $\rho_4(x_3)$ is the eigenvalues distribution of the matrix $X_3$ in the matrix phase for the bosonic theory for both values $\tilde{\beta}=0$ and $\tilde{\beta}=2/9$.

Another non trivial check for this important conjecture is the theoretical prediction of the radius
  \begin{eqnarray}
<\frac{1}{N}Tr X_a^2>
&=&\frac{R^2}{2}.
\end{eqnarray}
This means that in the matrix phase the order parameter $<\frac{1}{N}Tr X_a^2>$, which is related to the radius, is  constant. Indeed, this is what we see on figures (\ref{radiusF1}) and (\ref{radiusF2}) with a good agreement between the Monte Carlo measurement and the theoretical prediction. The observed value of $<\frac{1}{N}Tr X_a^2>$ is slightly below the theoretical  prediction for $\tilde{\beta}=2/9$ throughout.  This is not the case for $\tilde{\beta}=0$ where the Monte Carlo measurement starts slightly below $R^2/2$ and then rises above it as we approach the transition to the fuzzy sphere. 

This can  potentially be a serious difference between the tow cases $\tilde{\beta}=2/9$ and $\tilde{\beta}=0$. The transition to the sphere in the case of $\tilde{\beta}=2/9$ is in the form of an abrupt jump but in the case of $\tilde{\beta}=0$ there is a slow rise in the matrix phase as we increase $\tilde{\alpha}$  before the actual jump.

The main conclusion of these successful measurements is the fact that the matrices $X_{\mu}$ in the matrix phase are commuting and thus they are diagonalizable with a joint eigenvalues distribution which is uniform inside a ball of dimension $R$.

\begin{figure}[htbp]
\begin{center}
\includegraphics[width=10.0cm,angle=-90]{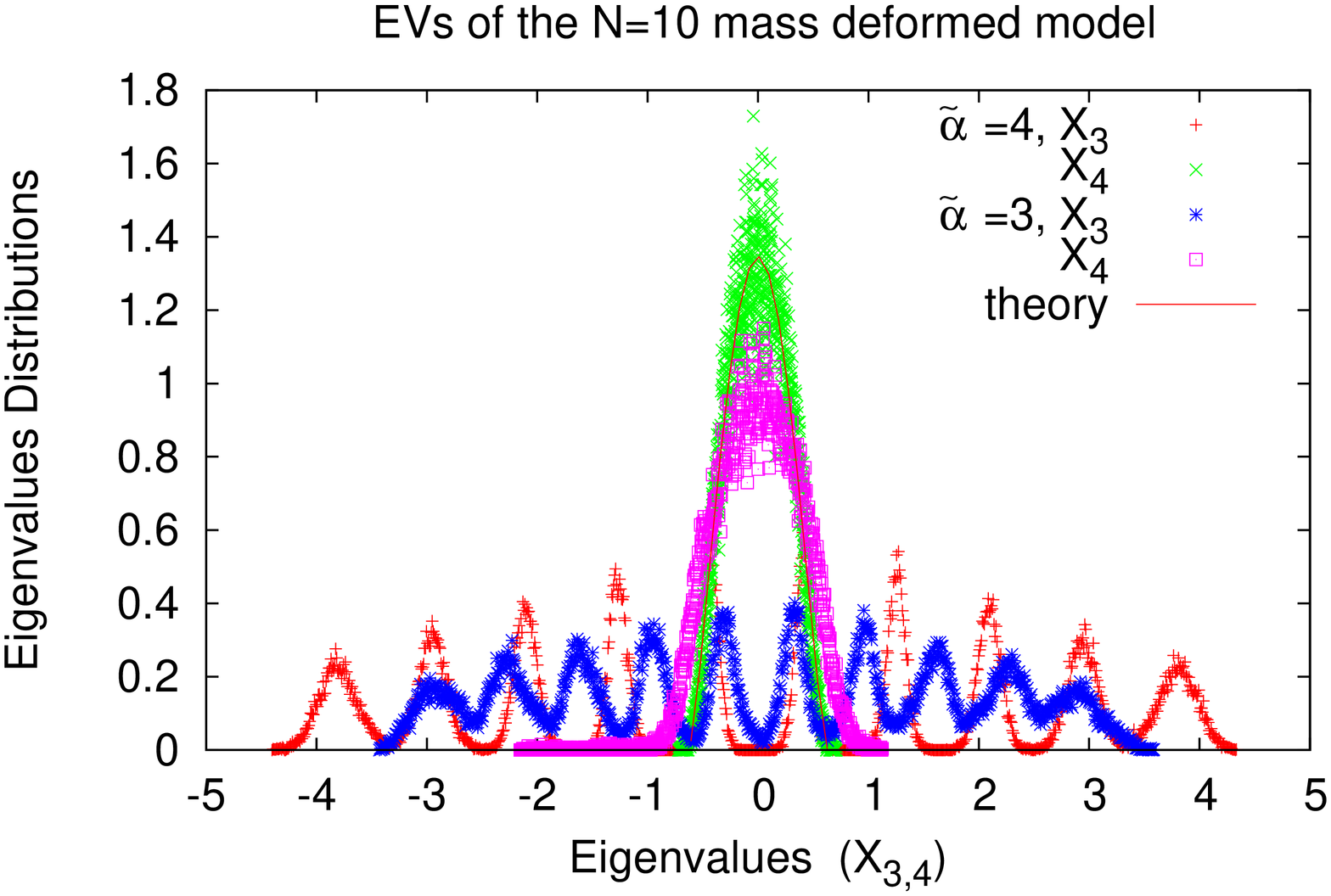}
\includegraphics[width=10.0cm,angle=-90]{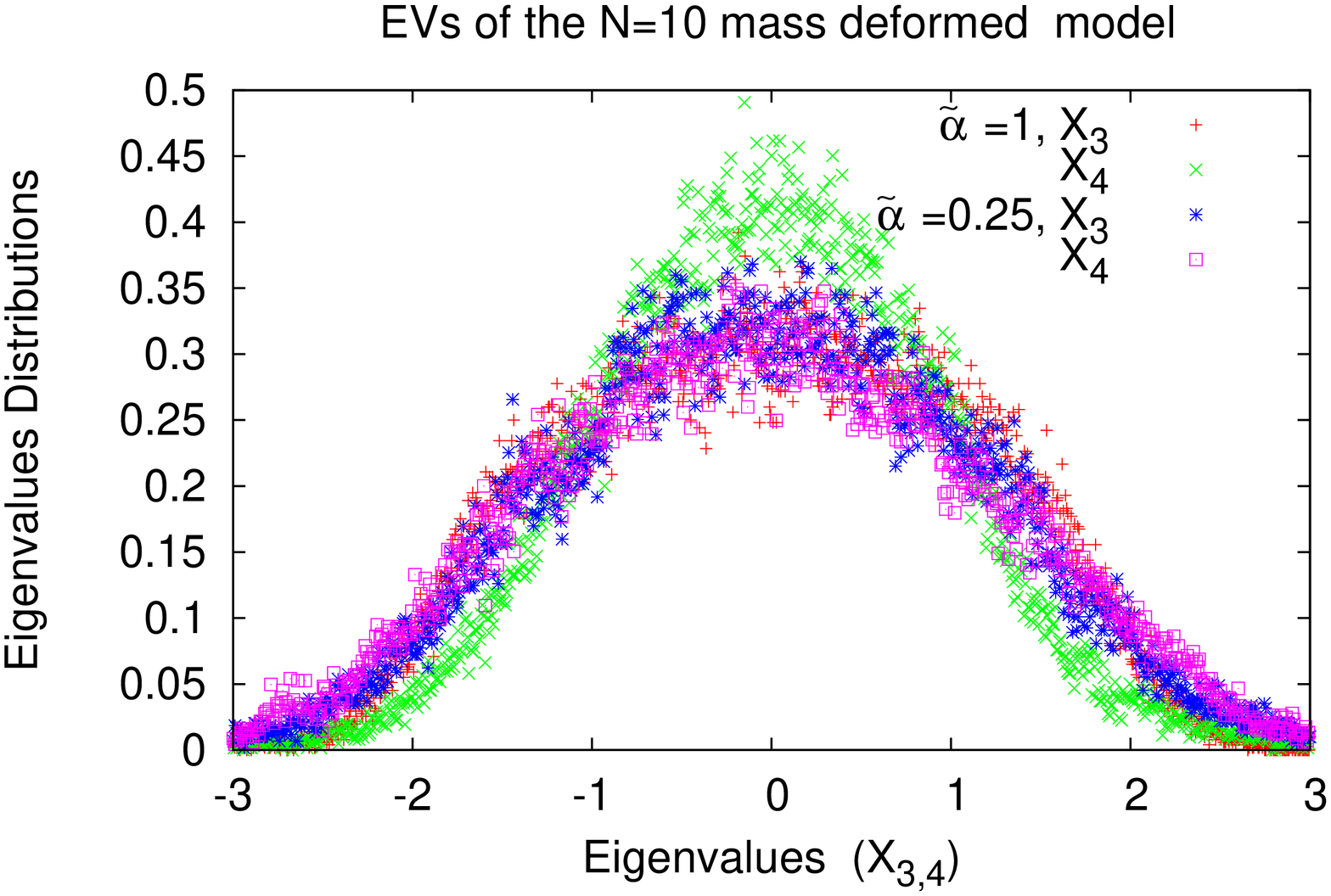}
\caption{The eigenvalues distributions of the mass deformed model. }\label{distrF1}
\end{center}
\end{figure}

\paragraph{Supersymmetric Theory} In the supersymmetric case we found it much easier to compute eigenvalues distributions with the value $\tilde{\beta}=2/9$ and thus we will only consider here the mass deformed model. A sample of the eigenvalues distributions of the mass deformed model is shown on figures (\ref{distrF1}) and (\ref{distrF2}). 

Again it is observed that the eigenvalues distribution of $X_3$ in the fuzzy sphere phase is given by an $N-$cut distribution corresponding to the $N$ eigenvalues $-(N-1)/2,...,(N-1)/2$ whereas in the matrix phase  the  eigenvalues distributions of $X_3$ is given by $\rho_4(x_3)$ with a much larger value of $R$ given for $\tilde{\alpha}=0.25$ by 
\begin{eqnarray}
R=2.851\pm 0.009. 
\end{eqnarray}
The eigenvalues distribution of $X_4$ is always centered around $0$ given by $\rho_4(x_4)$ with a value of $R$ which depends on the coupling constant $\tilde{\alpha}$. This  eigenvalues distribution coincides with the eigenvalues distribution of $X_3$ deep inside the matrix phase below $\tilde{\alpha}=0.25$.

The eigenvalues distribution $\rho_4$ is therefore universal in the sense that it describes the behavior of the eigenvalues of the matrix $X_4$ for all values of $\tilde{\alpha}$ and all values of $\tilde{\beta}$ and the behavior of the eigenvalues of the matrix $X_3$ in the matrix phase for all values of $\tilde{\beta}$. We note here the difference between this eigenvalues distribution $\rho_4$  and the eigenvalues distribution of the supersymmetric model with $\tilde{\alpha}=\tilde{\beta}=0$ \cite{Krauth:1999qw}. In the latter case the distribution extends from $-\infty $ to $+\infty $ and goes as $1/x^3$ for large eigenvalues. It is not clear to us at this stage how the two distributions relate to each other.

The point at which the eigenvalues distributions of $X_3$ and $X_4$ coincide may be taken as the measure for the crossover transition point and thus for $N=10$ this occurs at a point between $\tilde{\alpha}=1$ and $\tilde{\alpha}=0.25$.
\subsection{Remarks: $D=3$ Yang-Mills Matrix Models and Scalar Fluctuations}
The $D=3$ Yang-Mills matrix models we can immediately consider here can be obtained from the above $D=4$ models by simply setting the fourth matrix $X_4$ to zero.  This is different from the IKKT supersymmetric Yang-Mills matrix model in $D=3$ by the fact that it involves a determinant instead of a Pffafian and  as a consequence the path integrals of these theories are convergent. 

The physics of these $D=3$ models is identical to the physics of the $D=4$ models in the sense that there is a first/second order phase transition from a background geometry (the fuzzy sphere) to commuting matrices which in the presence of dynamical fermions is turned into a slow crossover transition. The most important difference is the natural expectation that the eigenvalues distributions of the matrices $X_a$ in the matrix phase  must be distributed according to the formula

\begin{eqnarray}
\rho_3(x_3)=\frac{3}{4R^3}(R^2-x_3^2).\label{3ball}
\end{eqnarray}
By analogy with the $D=4$ formula (\ref{4ball}) this distribution can be derived from the assumption that the joint eigenvalues distribution of the $3$ matrices $X_1$, $X_2$, $X_3$ is uniform inside a three dimensional ball of radius $R$. In the next section we will also derive this distribution for the $D=4$ Yang-Mills matrix model with a particular choice of the parameters with $\beta_4\neq 0$ (see (\ref{model0})). 

\begin{figure}[htbp]
\begin{center}
\includegraphics[width=10.0cm,angle=-90]{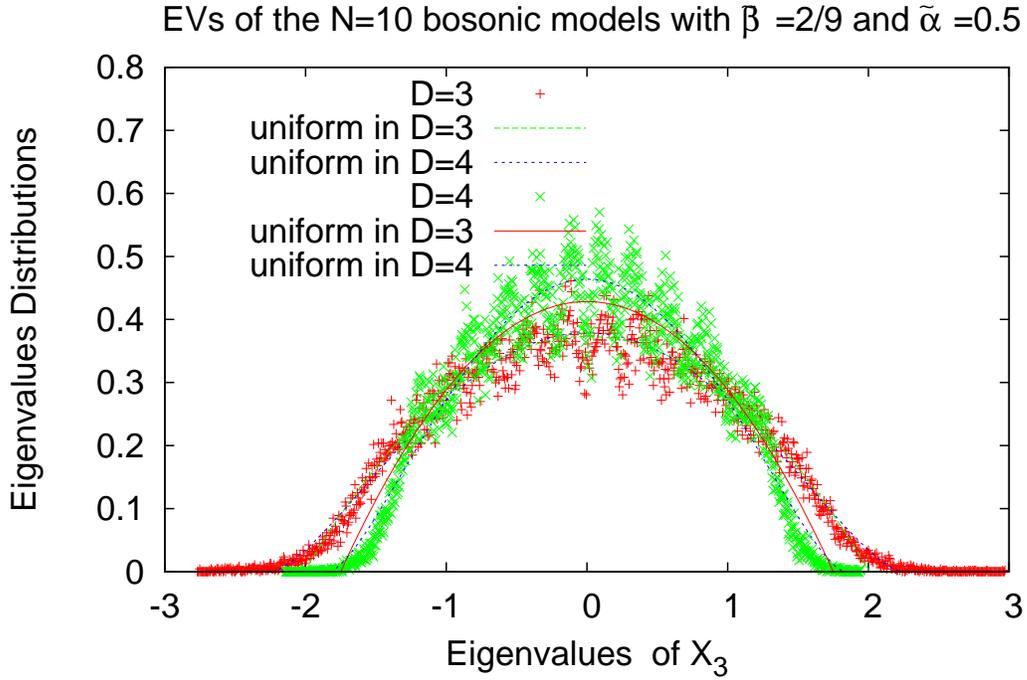}
\includegraphics[width=10.0cm,angle=-90]{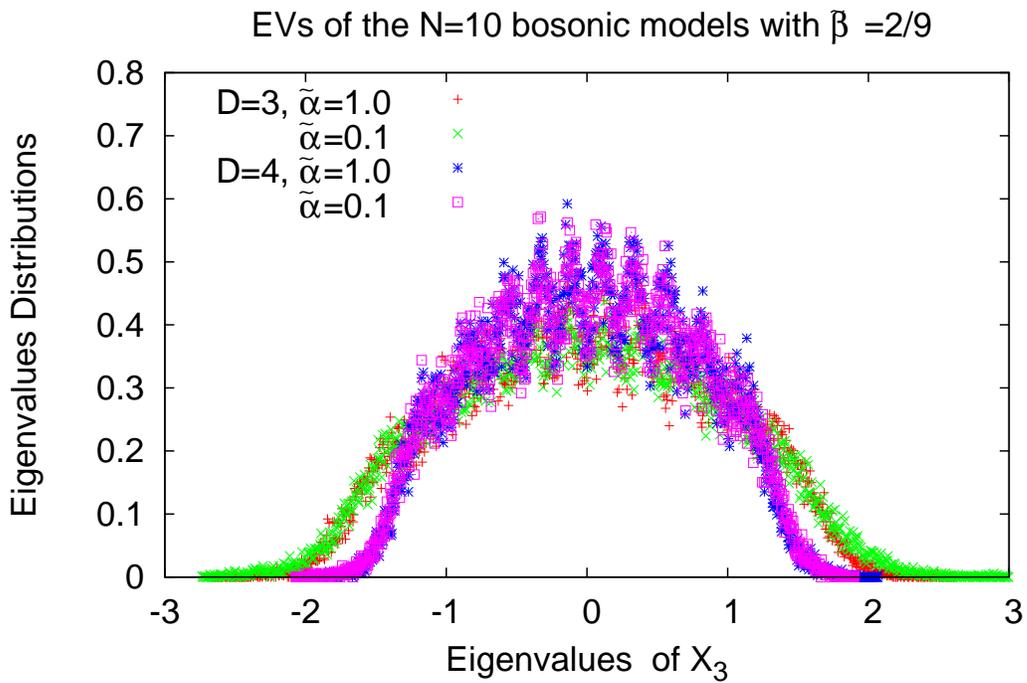}
\caption{The eigenvalues distributions of $X_3$ of the $=3,4$ bosonic models with $\tilde{\beta}=2/9$ in the matrix phase. }\label{D3D4}
\end{center}
\end{figure}

Monte Carlo measurment of the radius gives the value

\begin{eqnarray}
R\simeq 2.
\end{eqnarray}
As shown on the first graph of the figure (\ref{D3D4}) a sample of the data for the $N=10$ three-dimensional bosonic Yang-Mills matrix model with $\tilde{\beta}=2/9$ and $\tilde{\alpha}=0.5$ is shown. Clearly it can be fit nicely to (\ref{3ball}) with $R=2$. In performing the fitting in three dimensions we have to cut the tails in order to get a sensible answer.  As it turns out the same three dimensional data can also be fit to the four dimensional  prediction (\ref{4ball}).  We also show the data for the $N=10$ four-dimensional bosonic Yang-Mills matrix model with $\tilde{\beta}=2/9$ and $\tilde{\alpha}=0.5$ for comparison.

In the Monte Carlo data of the $D=3$ Yang-Mills matrix models reported in this article it was not possible to resolve the ambiguity between the two fits (\ref{4ball}) and (\ref{3ball}). However high precision runs performed in \cite{rodrigo} seems to indicate that indeed the three dimensional prediction  (\ref{3ball}) is the correct behavior for the eigenvalues distributions of the $D=3$ Yang-Mills matrix models.

The second remark we would like to discuss in this section concerns the dependence on $N$ and $\tilde{\alpha}$ of the eigenvalues distributions  $\rho_4$ and $\rho_3$ given in (\ref{4ball}) and (\ref{3ball}) respectively. As shown on the second graph of the figure (\ref{D3D4}) the distributions $\rho_4$ and $\rho_3$ are independent of $\tilde{\alpha}$. Similarly we can show that these distributions are independent of $N$.

The third remark concerns the eigenvalues distributions of the normal scalar field in the fuzzy sphere phase which is define by $\phi=(X_a^2-\phi^2c_2)/(2\phi\sqrt{c_2})$ \cite{CastroVillarreal:2004vh}. The behavior in both the $D=3$ and $D=4$ models is the same although we have to note that the effective values $c_{2,\rm eff}$ in $D=3$ and $D=4$ are slightly different. This is quite natural as the three dimensional model is more stable in the sense that it has a lower critical value $\tilde{\alpha}_*$. The central observation here is that we can nicely fit these eigenvalues distributions to the eigenvalues distribution  $\rho_4$ given in (\ref{4ball}).

\section{Summary and Future Directions}
In this article we employed  
the Monte Carlo method to study nonperturbatively Yang-Mills matrix models in $D=4$ with mass terms. We can summarize the main results, findings and conjectures of this work as follows:
\begin{itemize}
\item{}By imposing the requirement of supersymmetry and $SO(3)$ covariance we have shown that there exists a single mass deformed Yang-Mills quantum mechanics in $D=4$ which preserves all four real supersymmetries of the original theory although in a deformed form. This is the $4$ dimensional analogue of the $10$ dimensional BMN model. Full reduction yields a unique mass deformed $D=4$ Yang-Mills matrix model. This latter $4$ dimensional model is the analogue of the $10$ dimensional IKKT model.
\item{} By using cohomological deformation of supersymmetry we constructed a one-parameter ($\zeta_0$) family of cohomologically  deformed  $D=4$ Yang-Mills matrix models which preserve two supercharges. The mass deformed model is one limit ($\zeta_0\longrightarrow 0$) of this 
one-parameter family of cohomologically  deformed  Yang-Mills models.
\item{}We studied the models with the values $\tilde{\beta}=0$ and $\tilde{\beta}=2/9$ where $\tilde{\beta}$ is the mass parameter of the bosonic matrices $X_a$. The second model is special in the sense that classically the configurations $X_a\sim L_a,X_4=0$ is degenerate with the configuration $X_a=0,X_4=0$.
\item{}The Monte Carlo simulation of the bosonic  $D=4$ Yang-Mills matrix model with mass terms shows the existence of an exotic first/second order transition from a phase with a well defined background geometry given by the famous fuzzy sphere to a phase with commuting matrices with no geometry in the sense of Connes. The transition looks first order due to the jump in the action whereas it looks second order due to the divergent peak in the specific heat.
\item{}The fuzzy sphere is less stable as we increase the mass term of the bosonic matrices $X_a$, i.e. as we increase $\tilde{\beta}$. For $\tilde{\beta}=2/9$ we find the critical value $\tilde{\alpha}_*=4.9$ whereas for $\tilde{\beta}=0$ we find the critical value $\tilde{\beta}=2.55$. 
\item{}The measured critical line in the plane $\tilde{\alpha}-\tilde{\beta}$ agrees well with the theoretical prediction coming from the effective potential calculation.  
\item{}The order parameter of the transition is given by the inverse radius of the sphere defined by $1/r=Tr X_a^2/(\tilde{\alpha}^2c_2)$. The radius is equal to $1/\phi^2$ (where $\phi$ is the classical configuration) in the fuzzy sphere phase. At the transition point the sphere expands abruptly to infinite size. Then as we decrease the inverse temperature (the inverse gauge coupling constant) $\tilde{\alpha}$, the size of the sphere shrinks fast to $0$, i.e. the sphere evaporates.

\item{}The fermion determinant is positive definite for all gauge configurations in $D=4$. We have conjectured that the path integral is convergent as long as the scalar curvature (the mass of the fermionic matrices) is zero.  
\item{}We have simulated the two models $\tilde{\beta}=0$ and $\tilde{\beta}=2/9$ with dynamical fermions. The model with  $\tilde{\beta}=0$ has two supercharges while the model  $\tilde{\beta}=2/9$ has a softly broken supersymmetry since in this case we needed to set by hand the scalar curvature to zero in order to regularize the path integral.

Thus $\tilde{\beta}=0$ is amongst the very few models (which we known of) with exact supersymmetry which can be probed and accessed with the Monte Carlo method.

\item{}The fuzzy sphere is stable for the supersymmetric $D=4$ Yang-Mills matrix model with mass terms in the sense that the bosonic phase transition is turned into a very slow crossover transition. The transition point $\tilde{\alpha}$ is found to scale to zero with $N$. There is no jump in the action nor a peak in the specific heat.

\item{}The fuzzy sphere is stable also in the sense that the radius is equal $1/\phi^2$ over a much larger region then it starts to decrease slowly as we decrease   the inverse temperature $\tilde{\alpha}$ until it reaches $0$ at $\tilde{\alpha}=0$. We claim that the value where the radius starts decreasing becomes smaller as we increase $N$. 

The model at $\tilde{\alpha}=0$ can never sustain the geometry of the fuzzy sphere since it is the non deformed model so in some sense the transition to commuting matrices always occurs and in the limit $N\longrightarrow \infty$ it will occur at  $\tilde{\alpha}_*\longrightarrow 0$. 
 
\item{}We have spent a lot of time in trying to determine the eigenvalues distributions of the matrices $X_{\mu}$ in both the bosonic and supersymmetric theories. A universal behavior seems to emerge with many subtleties. These can be summarized as follows:
\begin{itemize}
\item{}In the fuzzy sphere the matrices $X_a$ are given by the $SU(2)$ irreducible representations $L_a$. For example diagonalizing the matrix $X_3$ gives $N$ eigenvalues between $(N-1)/2$ and $-(N-1)/2$ with a step equal $1$, viz $m=(N-1)/2,(N-3)/2,...,-(N-3)/2,-(N-1)/2$.

\item{}In the matrix phase the matrices $X_{\mu}$ become commuting. More explicitly the eigenvalues distribution of any of the matrices $X_{\mu}$ in the matrix phase is given by the non-polynomial law 
\begin{eqnarray}
\rho_4(x)=\frac{8}{3\pi R^4}(R^2-x^2)^{\frac{3}{2}}.
\end{eqnarray}
This can be obtained from the conjecture that the joint probability distribution of the four matrices $X_{\mu}$ is uniform inside a solid ball with radius $R$. 
\item{}In the matrix phase the eigenvalues distribution of any of the $X_a$, say $X_3$,  is given by the above  non-polynomial law with a radius $R$ independent of $\tilde{\alpha}$ and $N$.

\item{}This  is also confirmed by computing the radius in this distribution and comparing to the Monte Carlo data. 

\item{}A very precise measurement of the transition point can be made by observing the point at which the eigenvalues distribution of $X_3$ undergoes the transition from the $N$-cut distribution to the above  non-polynomial law.

\item{}The eigenvalues distribution of $X_4$  is always given by the above  non-polynomial law, i.e. for all values of $\tilde{\alpha}$, with a radius $R$ which depends on $\tilde{\alpha}$ and $N$.

\item{}Another signal that the matrix phase is fully reached is when the eigenvalues distribution of $X_4$ coincides with that of $X_3$. From this point downward the eigenvalues distribution of $X_4$ ceases to depend on $\tilde{\alpha}$ and $N$. 

\item{}Monte Carlo measurements seems to indicate that $R=1.8$ for bosonic models and $R=2.8$ for supersymmetric models. The distribution becomes wider in the supersymmetric case.
\item{}We have also observed that the eigenvalues of the normal scalar field $X_a^2-c_2$ in the fuzzy sphere are also distributed according to the above non-polynomial law. This led us to the conjecture that the eigenvalues of the gauge field on the background geometry are also distributed according to the above non-polynomial law. Recall that the normal scalar field is the normal component of the gauge field to the background geometry which is the sphere here.
\end{itemize}
\item{}In the $D=3$ Yang-Mills matrix model with mass terms the eigenvalues distribution becomes polyonomial (parabolic) given by
 \begin{eqnarray}
\rho_3(x)=\frac{3}{4R^3}(R^2-x^2).
\end{eqnarray}
It was difficult for us in this article to differentiate with certainty between the two distributions $\rho_4$ and $\rho_3$ in the three dimensional setting.
\item{}Finally, we conjecture that the transition from a background geometry to the phase of commuting matrices is associated with spontaneous supersymmetry breaking. Indeed mass deformed supersymmetry preserves the fuzzy sphere configuration but not diagonal matrices.
\end{itemize}

Among the future directions that can be considered we will simply mention the following four points:
\begin{itemize} 
\item{}Higher precision Monte Carlo simulations of the models studied in this article is the first obvious direction for future investigation. The most urgent question (in our view) is the precise determination of the behavior of the eigenvalues distributions in $D=4$ and $D=3$. An analytical derivation of $\rho_3$ and especially $\rho_4$ is an outstanding problem.
\item{}Finding matrix models with emergent $4$ dimensional background geometry is also an outstanding problem. 
\item{}Models for emergent time, and to a lesser extent emergent gravity, and as a consequence emergent cosmology are very rare.
\item{}Monte Carlo simulation of supersymmetry based on matrix models seems to be a very promising goal.
\item{}A complete analytical understanding of the emergent geometry transition observed in Yang-Mills matrix models with mass terms is also an   outstanding problem.

In \cite{ydri_preparation}, we have attempted to compute the above eigenvalues distributions analytically. Using localization techniques we were able to find a special set of parameters for which the $D=4$ Yang-Mills matrix model with mass terms can be reduced to the three dimensional Chern-Simons (CS) matrix model. The saddle-point method leads then immediately to the eigenvalues distributions $\rho_3$. We believe that our theoretical prediction for the value of $R$ is reasonable compared to the Monte Carlo value.  We have also made a preliminary comparison between the dependence of $R$ on $\alpha$ in the hermitian and antihermitian CS matrix models. The hermitian case seems more appropriate for the description of the eigenvalues of $X_3$ whereas the antihermitian case may be relevant to the description of the eigenvalues of $X_4$.
\end{itemize}

\paragraph{Acknowledgments:} I would like to thank Denjoe O'Connor for extensive discussions at various  stages of this project. I would like also to thank  R.Delgadillo-Blando and Adel Bouchareb for their collaboration. The numerical simulations reported in this article were conducted on the clusters of the Dublin Institute for Advanced Studies.

This research was partially supported by ``The National Agency for the Development of University Research (ANDRU)'' under PNR contract number  U23/Av58 (8/u23/2723).

\newpage

\end{document}